\documentclass[onecolumn, draftclsnofoot, 12pt]{IEEEtran}
\usepackage{amsmath}
\usepackage{cite}
\usepackage{amssymb}
\usepackage{amsfonts}
\usepackage{dsfont}
\usepackage{graphicx}
\usepackage{epsfig}
\usepackage{subfigure}
\usepackage{psfrag}
\usepackage{xcolor}
\usepackage{url}
\usepackage[colorlinks,linkcolor=black,urlcolor=black,anchorcolor=black,citecolor=black,hyperfootnotes=true]{hyperref}
\usepackage[ruled,vlined]{algorithm2e}

\newtheorem{corollary}{Corollary}

\newtheorem{lemma}{Lemma}
\newtheorem{theorem}{Theorem}

\newcommand{\mv}[1]{\mbox{\boldmath{$ #1 $}}}
\newtheorem{assumption}{Assumption}

\title{Massive MIMO Communication with Intelligent Reflecting Surface}
\author{Zhaorui Wang, Liang Liu, Shuowen Zhang, and Shuguang Cui
\thanks{Z. Wang, L. Liu, and S. Zhang are with the Department of Electronic and Information Engineering, The Hong Kong Polytechnic University, Hong Kong, China (e-mails:  zrwang2009@gmail.com, \{liang-eie.liu,shuowen.zhang\}@polyu.edu.hk).}
\thanks{S. Cui is with the Future Network of Intelligence
Institute, The School of Science and Engineering, the Chinese University of Hong Kong, Shenzhen, China (e-mail: shuguangcui@cuhk.edu.cn).}}
\begin{document}
\maketitle \thispagestyle{empty} \vspace{-0.3in}

\begin{abstract}
This paper studies the feasibility of deploying intelligent reflecting surfaces (IRSs) in massive MIMO (multiple-input multiple-output) systems to improve the performance of users in the service dead zone. One question of paramount importance is as follows: if the overhead of channel training and the computational complexity of algorithm design arising from the huge number of IRS reflecting elements and base station (BS) antennas have to be controlled, can we provide reasonable performance to the users with weak direct channels? This paper provides an affirm answer to this question. Specifically, to reduce the channel training overhead, we advocate a novel protocol for the uplink communication in the IRS-assisted massive MIMO systems. Under this protocol, the IRS reflection coefficients are optimized based on the channel covariance matrices, which are generally fixed for many coherence blocks, to boost the long-term performance. Then, given the IRS reflecting coefficients, the BS beamforming vectors are designed in each coherence block based on the effective channel of each user, which is the superposition of its direct and reflected user-IRS-BS channels, to improve the instantaneous performance. Since merely the user effective channels are estimated in each coherence block, the training overhead of this protocol is the same as that in the legacy wireless systems without IRSs. Moreover, in the asymptotic regime that the numbers of IRS elements and BS antennas both go to infinity with a fixed ratio, we manage to first characterize the minimum mean-squared error (MMSE) estimators of the user effective channels and then quantify the closed-form user achievable rates as functions of channel covariance matrices with channel training overhead and estimation error taken into account. Interestingly, it is shown that the properties of channel hardening and favorable propagation still hold for the user effective channels, and satisfactory user rates are thus achievable even if simple BS beamforming solutions, e.g., maximal-ratio combining, are employed. Finally, thanks to the rate characterization, we design a low-complexity algorithm to optimize the IRS reflection coefficients based on channel covariance matrices.
\end{abstract}

\section{Introduction}\label{sec:Introduction}
\subsection{Motivation}
It is well accepted that massive MIMO (multiple-input multiple-output) will be a mainstream feature for enhanced mobile broadband communication in the fifth-generation (5G) and beyond 5G networks. After the seminal work\cite{marzetta2010noncooperative}, both solid theory and practical solutions have been devised for massive MIMO communication over the past decade, covering spectrum efficiency analysis\cite{ngo2013energy,hoydis2013massive,yang2013performance,marzetta2016fundamentals,bjornson2014massive,bjornson2015optimal}, pilot contamination and decontamination\cite{ngo2011analysis,Muller2014MIMO}, etc. However, despite the appealing properties of channel hardening, i.e., the strength of user channels does not fade over time, and favorable propagation, i.e., user channels are asymptotically orthogonal with each other, brought by massive MIMO, it remains an open problem about how to guarantee the performance of the users in the service dead zone, e.g., indoor users with thick walls between them and the base station (BS) or outdoor users surrounded by many tall buildings, when the severe channel attenuation cannot be compensated by the channel hardening gain.

\begin{figure}[t]
	\centering
	\includegraphics[width=10cm]{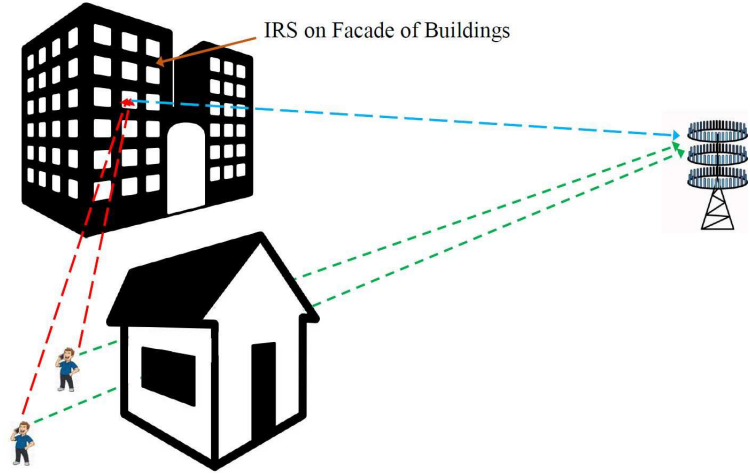}\vspace{-10pt}
	\caption{An IRS-assisted multiuser massive MIMO communication system.}\label{Fig1}\vspace{-15pt}
\end{figure}

In this paper, we study the feasibility of deploying intelligent reflecting surfaces (IRSs) in the massive MIMO communication systems for maintaining the performance of the users in the service dead zone. IRS is a planar surface consisting of a vast number of reflecting elements. By inducing phase shift to the incident signal at each reflecting element, the IRS is able to modify the channels between the BS and the users to be more favorable for communication\cite{Liaskos08,Renzo19,Basar19}. Moreover, the IRS is usually of thin material, allowing it to be easily attached to the ceilings and walls in indoor environments, and coated on the building facade in outdoor environments. Motivated by the above advantages, it is a natural idea to deploy IRSs for providing alternative user-IRS-BS communication links to the users within the service dead zone, as shown in Fig. \ref{Fig1}.

In the literature, a great amount of endeavor has been devoted to the joint optimization of the BS beamforming vectors and the IRS reflecting coefficients when the BS is equipped with a small or moderate number of antennas, and it is shown that the user signal-to-interference-plus-noise ratio (SINR) can be greatly improved by IRS\cite{wu2019intelligent,zhang2020capacity,Huang2019IRS,Shuowen21,Panrelay,PanSWIPT,Guo2020IRS,Yang2020IRS,Yu2020IRS,Hou2020IRS}. However, the above results are not convincing to verify the effectiveness of IRSs for serving the users at the service dead zone in a massive MIMO system, due to the following reasons. First, in an IRS-assisted massive MIMO system, the number of coefficients in the user-BS channels and the user-IRS-BS channels is huge. As a result, reducing the channel estimation overhead is a challenging issue before we plan to reap the beamforming gain brought by the IRS. Second, a joint design of the BS beamforming vectors and the IRS reflecting coefficients at each coherence block is intractable in IRS-assisted massive MIMO systems, considering the vast number of optimization variables associated with the BS antennas and IRS reflecting elements. In this regard, performance analysis given simple BS/IRS strategies and in the asymptotic regime where both the numbers of BS antennas and IRS reflecting elements go to infinity is more relevant than sophisticated optimization. Last, although the well-known channel hardening and favorable propagation properties hold in the conventional massive MIMO systems, it is not clear whether these beneficial properties are still true for the user-IRS-BS channels in IRS-assisted massive MIMO systems such that the users can rely on the reflecting channels when their direct channels are weak. In this paper, we aim to show that under certain novel protocol designed for the IRS-assisted massive MIMO systems, the asymptotic user rates are reasonable even under simple BS/IRS design because the channel training overhead can be significantly reduced, and the channel hardening as well as favorable propagation properties hold for the user-IRS-BS channels.

\subsection{Prior Works}
The pioneering works  \cite{marzetta2010noncooperative,bjornson2014massive,yang2013performance,marzetta2016fundamentals,ngo2013energy,hoydis2013massive} showed that the massive MIMO technology can significantly improve the throughput over the conventional MIMO technology even if the channel estimation overhead is considered. However, there still exists the service dead zones in massive MIMO systems where the direct channel is relatively weak due to blockage. To solve this problem, several works have investigated the possibility of deploying IRSs in massive MIMO systems to serve the users in the service dead zone. Specifically, in the asymptotic regime where the number of BS antennas is infinite but the number of IRS elements is finite, \cite{mass_irs} characterized the user achievable rate in an IRS-assisted system where the channel estimation overhead is also taken into consideration. The main conclusion in this regime is that the favorable propagation property no longer holds for user channels, and the zero-forcing (ZF) receive beamforming, instead of the maximal-ratio combining (MRC) beamforming, should be used at the BS. However, in a practical IRS-assisted massive MIMO system, the number of IRS elements may be much larger than that of the BS antennas. It is thus crucial to focus on another asymptotic regime where both the numbers of the IRS elements and the BS antennas go to infinity, and study whether the favorable propagation property holds. In addition, \cite{li2019joint,jamali2019intelligent} devised the algorithms to design the beamforming vectors at the BS and the reflection coefficients at the IRS based on the instantaneous channels. However, the overhead for channel estimation, which is huge in IRS-assisted communication systems \cite{wang2020channel,wang2019channel_cof,elbir2020deep}, and computational complexity to optimize so many variables are not considered in these works. This motivates us to focus on new mechanism for IRS-assisted communication with low channel training overhead and implementation complexity.

\subsection{Main Contributions}
In this paper, we consider the uplink communication in IRS-assisted massive MIMO systems, in which the users in the service dead zone can rely on the alternative user-IRS-BS channels provided by the IRSs for communication. In such systems, the BS may have tens or hundreds of antennas, while the IRS may have hundreds or thousands of reflecting elements. This motivates us to focus on the asymptotic regime where the numbers of BS antennas and IRS reflection elements both go to infinity, but with a fixed ratio. In this regime, we aim to characterize the user achievable rates when the channel training overhead and estimation error are taken into consideration. The main contributions of this paper are as follows.

First, we advocate a novel two-timescale communication protocol \cite{Jin19,zhao2020intelligent,kammoun2020asymptotic,papazafeiropoulos2021intelligent,jia2020analysis} for our considered IRS-assisted massive MIMO systems to balance the channel estimation overhead, the optimization complexity, and the IRS beamforming gain. Under this protocol, the IRS reflecting coefficients are designed to optimize the long-term performance based on the statistical channel state information (CSI), i.e., channel covariance matrices, which may keep approximately constant for a large number of channel coherence blocks.  \footnote{In practice, we can first estimate the user-IRS channels and the IRS-BS channels separately based on the methods proposed in \cite{liu2020matrix,hu2021two}, and then estimate the corresponding channel covariance matrices based on the techniques proposed in \cite{liang2001downlink,Chen10}.}. As a result, the IRS optimization is conducted at a relatively low frequency. On the other hand, given the IRS reflection coefficients, the BS beamforming vectors are designed at each coherence block to optimize the short-term performance based on the instantaneous CSI of each user's effective channel, which is the superposition of its direct user-BS channel and reflected user-IRS-BS channels. Note that the dimension of each user's effective channel equals to the number of BS antennas, similar to the conventional massive MIMO systems without IRS. As a result, under this protocol, the overhead for channel estimation at each coherence block is independent of the number of IRS elements. This is in sharp contrast to the existing works \cite{wu2019intelligent,zhang2020capacity,Huang2019IRS,Guo2020IRS,Yang2020IRS,Yu2020IRS,Hou2020IRS} where the IRS reflection coefficients and BS beamforming vectors are jointly optimized based on the users' direct channels as well as reflected channels via the IRS at each coherence block such that the channel estimation overhead is linear to the number of IRS elements \cite{wang2020channel}. Moreover, the considered protocol is also appealing from the computational complexity perspective, since the IRS reflection coefficients are optimized at a low frequency, while the existing beamforming design schemes can be directly applied by the BS given the user effective channels.

Second, under the above novel communication protocol and our interested asymptotic regime, we manage to characterize the closed-form (approximated) minimum mean-squared error (MMSE) estimators for estimating the user effective channels and the corresponding estimation mean-squared error (MSE) at each coherence block. In general, the distribution of the user effective channels is very complicated in IRS-assisted communication systems because the user-IRS-BS channels are products of the random user-IRS and IRS-BS channels. In this paper, we manage to express the user effective channels as the summation of $N+1$ independent channels, where $N$ denotes the number of IRS reflecting elements. Then, based on the multivariate Lindeberg-Feller central limit theorem \cite{van2000asymptotic}, it is shown that the distribution of user effective channels approaches the joint Gaussian distribution in our interested asymptotic regime. As a result, the classic estimation theory for estimating Gaussian channels are applied to design the MMSE channel estimators and quantify their estimation MSE in the IRS-assisted massive MIMO systems.

Third, under our interested communication protocol and asymptotic regime, we show that the user effective channels are independent of each other. As a result, the favorable propagation property in the conventional massive MIMO systems, i.e., the inner produce of two users' channels approaches zero when the number of BS antennas goes to infinity, still holds in the IRS-assisted massive MIMO systems under our considered asymptotic regime. Note that in another asymptotic regime where the number of BS antennas goes to infinity but the number of IRS reflecting elements is finite, our previous work \cite{mass_irs} showed that the property of favorable propagation does not hold because the identical IRS-BS channel appears in all the users' reflected channels such that the user effective channels are not independent. Thanks to the multivariate Lindeberg-Feller central limit theorem \cite{van2000asymptotic}, if the number of IRS elements goes to infinity as well, then user effective channels become orthogonal again in the massive MIMO systems.

Fourth, thanks to the favorable propagation property, we employ the MRC beamforming vectors at the BS in this work. Then, under our interested communication protocol and asymptotic regime, we characterize the closed-form user achievable rate expressions as functions of statistical CSI, i.e., channel covariance matrices, based on the random matrix theory \cite{couillet2011random}, where the channel estimation time duration and the channel estimation error are both considered. Note that the IRS-assisted massive MIMO communication was also considered in  \cite{zhi2021statistical,zhi2021ergodic,bjornson20}. However, the channel hardening and favorable propagation properties were not revealed in the above works, and the rate expression is thus in much more complicated form.

Last, thanks to the user rate characterization as functions of the channel covariance matrices, we provide a successive convex approximation technique \cite{marks1978general} to minimize the long-term user total transmit power subject to the individual user rate constraints. Numerical results are provided to verify the effectiveness of the proposed optimization algorithm.

The rest of this paper is organized as follows. Section \ref{sec:SYS} presents the system model for IRS-assisted massive MIMO communication. Section \ref{sec:protocol design} introduces a two-timescale communication protocol. Section \ref{sec:channel estimation} presents the distribution of the user effective channels and the corresponding MMSE channel estimators. Section \ref{sec:Date transmission} characterizes user achievable rates as functions of channel covariance matrices when channel estimation overhead and error are considered. Section \ref{sec:design phi} optimizes the IRS reflection coefficients based on the rate characterization. Finally, Section \ref{sec:Conclusions} concludes the paper.

\section{System Model}\label{sec:SYS}
We study a  massive MIMO system in which $K$ single-antenna users simultaneously communicate with a BS equipped with $M$ antennas in the uplink. We assume that the users are in the service dead zone where their direct communication channels to the BS are very weak with the presence of obstacles such as tall buildings. In such a scenario, an IRS equipped with $N$ reflecting elements is deployed to enhance the data transmission rates of these users, as shown in Fig. \ref{Fig1}. Specifically, the reflection elements of the IRS are able to dynamically adjust their reflection coefficients to re-scatter the electromagnetic waves from the users to the BS with focused energy. Let $\phi_{n}=\alpha_n e^{j\theta_n}$ denote the reflection coefficient of the $n$-th IRS element, $n=1,\ldots,N$, where $0<\alpha_n\le 1$ and $0\le\theta_n<2\pi$ denote the amplitude and phase, respectively.

We assume quasi-static block-fading channels, in which all channels remain approximately constant in each coherence block consisting of $T$ symbols. Let $\mv{h}_k\in\mathbb{C}^{M\times1}$ denote the direct channel from the $k$-th user to the BS,  $t_{k,n}\in\mathbb{C}$ denote the channel from the $k$-th user to the $n$-th IRS element, and  $\mv{r}_n\in\mathbb{C}^{M\times1}$ denote the channel from the $n$-th IRS element to the BS, $k=1,\ldots,K$, $n=1,\ldots,N$. In this paper, the direct channels $\mv{h}_k$'s are modeled as
\begin{align}
\mv{h}_k=(\mv{C}_k^{{\rm B}})^{\frac{1}{2}}\tilde{\mv{h}}_k, ~~\forall k,
\end{align}
where $\mv{C}_k^{{\rm B}}\in\mathbb{C}^{M\times M}\succ \mv{0}$ denotes the BS receive correlation matrix for user $k$, and  $\tilde{\mv{h}}_k\sim \mathcal{CN}(\mv{0},\beta_k^{{\rm BU}}\mv{I})$ follows the independent and identically distributed (i.i.d.) Rayleigh fading channel model, with $\beta_k^{{\rm BU}}$ denoting the path loss of $\mv{h}_k$. Due to the blockages between the BS and the users, $\beta_k^{{\rm BU}}$'s are assumed to be small in this paper. In the extreme case of no direct link between the BS and user $k$, we have $\beta_k^{{\rm BU}}=0$. Moreover, we assume that $\tilde{\mv{h}}_k$'s are independent over $k$. Next, define  $\mv{R}=[\mv{r}_1,\cdots,\mv{r}_N]$ as the overall channel from the IRS to the BS. Then, $\mv{R}$ is modeled as
\begin{align}\label{eq: R}
\mv{R}=(\mv{C}^{{\rm B}})^{\frac{1}{2}}\tilde{\mv{R}}(\mv{C}^{{\rm I}})^\frac{1}{2},
\end{align}
where $\mv{C}^{{\rm B}}\in \mathbb{C}^{M\times M}\succ\mv{0}$ denotes the BS receive correlation matrix for the IRS, and $\mv{C}^{{\rm I}}\in \mathbb{C}^{N\times N}\succ\mv{0}$ denotes the IRS transmit correlation matrix for the BS.  In addition, $\tilde{\mv{R}}\sim \mathcal{CN}(\mv{0},N\beta^{{\rm BI}}\mv{I})$ follows the i.i.d. Rayleigh fading channel model, where $\beta^{{\rm BI}}$
denotes the path loss of $\mv{R}$. Last, define $\mv{t}_k=[t_{k,1},\cdots,t_{k,N}]^T$ as the overall channel from user $k$ to the IRS, $\forall k$. They are modeled as
\begin{align}
\mv{t}_k=(\mv{C}_k^{{\rm I}})^{\frac{1}{2}}\tilde{\mv{t}}_k,~~\forall k,
\end{align}
where $\mv{C}_k^{{\rm I}}\in \mathbb{C}^{N\times N}\succ\mv{0}$ denotes the IRS receive correlation matrix for user $k$, and  $\tilde{\mv{t}}_k\sim \mathcal{CN}(\mv{0}, \beta_k^{{\rm IU}}\mv{I})$ follows the i.i.d. Rayleigh fading channel model, with $\beta_k^{{\rm IU}}$ denoting the path loss of $\mv{t}_k$. We assume that $\tilde{\mv{t}}_k$'s are independent over $k$. Note that the channels $\mv{h}_k$'s, $\mv{R}$, and $\mv{t}_k$'s merely remain constant within each coherence block, while the channel correlation matrices $\mv{C}^{\rm B}$, $\mv{C}^{\rm B}_k$'s, $\mv{C}^{\rm I}$, and $\mv{C}^{\rm I}_k$'s can keep constant within a large number of coherence blocks.

Under the considered model, the received signal at the BS is expressed as
\begin{align}
\mv{y}&=\sum_{k=1}^{K}\mv{h}_k\sqrt{p_k}s_k+\sum_{k=1}^{K}\sum_{n=1}^{N}\phi_n\mv{g}_{k,n}\sqrt{p_k}s_k+\mv{z}\nonumber\\
&=\sum_{k=1}^{K}\mv{c}_k\sqrt{p_k}s_k+\mv{z}
 \label{eq:rec_sig},
\end{align}
where $p_k$ denotes the transmit power of user $k$, $s_k\in \mathcal{CN}(0,1)$ denotes the transmit symbol of user $k$,  $\mv{z}\sim\mathcal{CN}\left(\mv{0},\sigma^2\mv{I}\right)$ denotes the additive white Gaussian noise (AWGN) at the BS,
\begin{align}
\mv{g}_{k,n}=t_{k,n}\mv{r}_n, ~~\forall k,n, \label{eq: g_kn}
\end{align}
is the effective reflecting channel from user $k$ to the BS through the IRS reflecting element $n$, and
\begin{align}
\mv{c}_k=\mv{h}_k+\sum_{n=1}^{N}\phi_{n}\mv{g}_{k,n}, ~~\forall k, \label{eq: c}
\end{align}
is the effective channel between user $k$ and the BS contributed by both the direct channel and the reflecting channels via the IRS. Note that, for any user $k$, the mean and covariance matrix of its effective channel are $\mathbb{E}[\mv{c}_k]=\mv{0}$ and
\begin{align}
\mv{V}_k&=\mathbb{E}[\mv{c}_k\mv{c}_k^H]=\beta_k^{\rm BU}\mv{C}_k^{{\rm B}}+\beta_k^{\rm IU}\mathbb{E}[(\mv{C}^{\rm B})^{\frac{1}{2}}\tilde{\mv{R}}\mv{D}_k(\mv{\phi})\tilde{\mv{R}}^H(\mv{C}^{\rm B})^{\frac{1}{2}}] \nonumber \\ & =\beta_k^{\rm BU}\mv{C}_k^{{\rm B}}+\beta^{\rm BI}\beta_k^{\rm IU}{\rm tr}\left(\mv{D}_k(\mv{\phi})\right)\mv{C}^{\rm B} \label{eq: V_k1} \\
& = \beta_k^{\rm BU}\mv{C}_k^{{\rm B}}+\beta^{\rm BI}\beta_k^{\rm IU}\mv{\phi}^H\bar{\mv{C}}_k\mv{\phi}\mv{C}^{\rm B}, \label{eq: V_k}
\end{align}
respectively, where $\mv{\phi}=[\phi_1,\dots,\phi_N]^T$, and
\begin{align}
& \mv{D}_k(\mv{\phi})=(\mv{C}^{\rm I})^{\frac{1}{2}}{\rm diag}(\mv{\phi})\mv{C}_k^{\rm I}\left({\rm diag}(\mv{\phi})\right)^H(\mv{C}^{\rm I})^{\frac{1}{2}}, \label{eq: D_k} \\
& \bar{\mv{C}}_k=\mv{C}^{\rm I}\circ(\mv{C}_k^{\rm I})^T,\label{eq: C_k}
\end{align}
with $\circ$ denoting Hadamard product, ${\rm diag}(\mv{\phi})$ denoting a diagonal matrix with  $\mv{\phi}$ being  diagonal elements. Note that (\ref{eq: V_k1}) can be obtained via the eigenvalue decomposition of $\mv{D}_k(\mv{\phi})$ and the fact that $\tilde{\mv{R}}\mv{U}\sim \mathcal{CN}(\mv{0},N\beta^{{\rm BI}}\mv{I})$ if $\mv{U}\in \mathbb{C}^{N\times N}$ is a unitary matrix. Moreover, $\bar{\mv{C}}_k\succ\mv{0}$, $\forall k$, according to the Schur product theorem \cite{bapat1997nonnegative}.

After receiving the signals from the users, the BS applies a beamforming vector $\mv{w}_k\in \mathbb{C}^{M\times 1}$ to decode $s_k$, $k=1,\dots,K$, i.e.,
\begin{align}
\mv{\tilde{y}}_k=\sum_{j=1}^K\mv{w}_k^H\mv{c}_j\sqrt{p_j}s_j\hspace{-2pt}+\hspace{-2pt}\mv{w}_k^H\mv{z}, ~~\forall k. \label{eq: y}
\end{align}
Then, assuming perfect CSI at the BS, the SINR for decoding $s_k$ is
\begin{align}
\gamma_k=\frac{p_k\left|\mv{w}_k^H\left(\mv{h}_k+\sum_{n=1}^{N}\phi_{n}\mv{g}_{k,n}\right)\right|^2}{\sum\limits_{j\ne k}p_j\left|\mv{w}_k^H\left(\mv{h}_j+\sum_{n=1}^{N}\phi_{n}\mv{g}_{j,n}\right)\right|^2\hspace{-2pt}+\sigma^2\mv{w}_k^H\mv{w}_k}=\frac{p_k\left|\mv{w}_k^H\mv{c}_k\right|^2}{\sum\limits_{j\ne k}p_j\left|\mv{w}_k^H\mv{c}_j\right|^2\hspace{-2pt}+\sigma^2\mv{w}_k^H\mv{w}_k}, ~~\forall k. \label{eq:Sys-A1.9}
\end{align}

\section{A Novel Two-Timescale Communication Protocol}\label{sec:protocol design}
It can be observed from \eqref{eq:Sys-A1.9} that to maximize the user SINRs, at the beginning of each coherent block, the BS should jointly design its receive beamforming vectors $\mv{w}_k$'s and the IRS's reflection coefficients $\phi_n$'s based on the user-BS channels $\mv{h}_k$'s and user-IRS-BS channels $\mv{g}_{k,n}$'s\cite{Wu18,wu2019intelligent,han2020cooperative}, and then send the reflection coefficients to the IRS controller via the gateway network in every coherence block \cite{wu2019towards}. However, several practical challenges arise from the above joint design. First, the estimation of $\mv{h}_k$'s and $\mv{g}_{k,n}$'s is time-consuming due to the large number of channel coefficients in $\mv{g}_{k,n}$'s. Second, the optimization of such a large number of IRS reflecting coefficients in each coherence block is intractable in practice. Note that the above two practical challenges both arise from the dynamic optimization of $\phi_n$'s in each coherence block. Specifically, if $\phi_n$'s are fixed over time, then the optimization of $\mv{w}_k$'s at the BS is of lower complexity and merely requires the information of the effective channels $\mv{c}_k$'s as shown in \eqref{eq: c}, which can be estimated by  $K$ pilot symbols in each coherence block\cite{Hassibi03}. However, such a strategy totally loses the dynamic optimization gain thanks to the  ``configurable'' feature of the IRS.

To reap the dynamic IRS beamforming gain at low training overhead and computational complexity, in this paper, we consider a two-timescale optimization protocol \cite{Jin19,zhao2020intelligent}, where $\phi_n$'s are optimized based on the channel covariance matrices at a lower frequency  but $\mv{w}_k$'s are optimized based on the instantaneous effective channels $\mv{c}_k$'s at a higher frequency. Specifically, define ``covariance interval'' as the maximum time duration during which the channel covariance matrices $\mv{C}^{\rm B}$, $\mv{C}^{\rm I}$, $\mv{C}_k^{\rm B}$'s and $\mv{C}_k^{\rm I}$'s remain constant. Then, in each covariance interval which may consist of a large number of channel coherence blocks, the IRS employs a fixed pattern of reflecting coefficients $\phi_n$'s, which are designed based on the channel correlation matrices (please refer to Section \ref{sec:design phi}) to improve the average performance of the whole covariance interval. Next, given the optimized IRS reflection coefficients over the whole covariance interval, at the beginning of each coherence block, the BS can estimate the effective channels of the users, i.e., $\mv{c}_k$'s, based on their pilot signals (please refer to  Section \ref{sec:channel estimation}). Then, all the users can transmit their data to the BS in the remaining time of each coherence block, which then designs beamforming vectors $\mv{w}_k$'s based on the estimation of $\mv{c}_k$'s to decode user messages (please refer to Section \ref{sec:Date transmission}) to improve the rate performance in each coherence block.

In a practical IRS-assisted massive MIMO system, the number of BS antennas and IRS elements are very large. As a result, in this paper, we study the performance of the two-timescale protocol in the asymptotic regime where both $N$ and $M$ go to infinity with a fixed ratio $q$, i.e.,
\begin{align}\label{eqn:regime}
N, M\rightarrow \infty ~ {\rm with} ~ \frac{N}{M}=q.
\end{align}
To facilitate the analysis in the asymptotic regime \eqref{eqn:regime}, we make the following assumptions.
\begin{assumption}\label{assumption1}
	The correlation matrices $\mv{C}^{\rm B}$, $\mv{C}^{\rm B}_k$'s, $\mv{C}^{\rm I}$, and $\mv{C}^{\rm I}_k$'s have uniformly bounded spectral norm.	
\end{assumption}

\begin{assumption}\label{assumption1.5}
	Let $\sigma_k^{{\min}}$ denote the minimum singular value of the matrix
	\begin{align}
		\mv{L}_k=(\mv{C}^{\rm I})^{\frac{1}{2}}{\rm diag}(\mv{\phi})(\mv{C}_k^{\rm I})^{\frac{1}{2}},~~~ k=1,\ldots,K. \label{eq:Lk}
	\end{align}
	For each $\mv{L}_k$, there exists $l_k$ that is independent of $N$ such that
	\begin{align}
		\sigma_k^{{\min}}> l_k>0, ~~~ k=1,\ldots,K. \label{eq:lk}
	\end{align}
\end{assumption}
\begin{assumption}\label{assumption2}
	Define
	\begin{align}
		\tilde{\mv{C}}_{k,j}=\frac{(\mv{C}_k^{\rm I})^{\frac{1}{2}}({\rm diag}(\mv{\phi}))^H(\mv{C}^{\rm I})^{\frac{1}{2}}(\tilde{\mv{R}})^H\mv{C}^{\rm B}\tilde{\mv{R}}(\mv{C}^{\rm I})^{\frac{1}{2}}{\rm diag}(\mv{\phi})(\mv{C}_j^{\rm I})^{\frac{1}{2}}}{N}, ~~\forall k,j, \label{eq: III-A0}
	\end{align}
	and
	\begin{align}
		\hat{\mv{C}}_{k,j}=\frac{(\mv{C}_k^{\rm B})^{\frac{1}{2}}(\mv{C}^{\rm B})^{\frac{1}{2}}\tilde{\mv{R}}(\mv{C}^{\rm I})^{\frac{1}{2}}{\rm diag}(\mv{\phi})(\mv{C}_j^{\rm I})^{\frac{1}{2}}}{N},~~\forall k,j. \label{eq: III-A01}
	\end{align}
	Then, $\tilde{\mv{C}}_{k,j}$ and $\hat{\mv{C}}_{k,j}$  have almost surely uniformly bounded
	spectral norm.
\end{assumption}

\begin{figure}[t]
	\begin{center}
		\subfigure[Validation under Covariance Model 1.]{\scalebox{0.58}{\includegraphics*{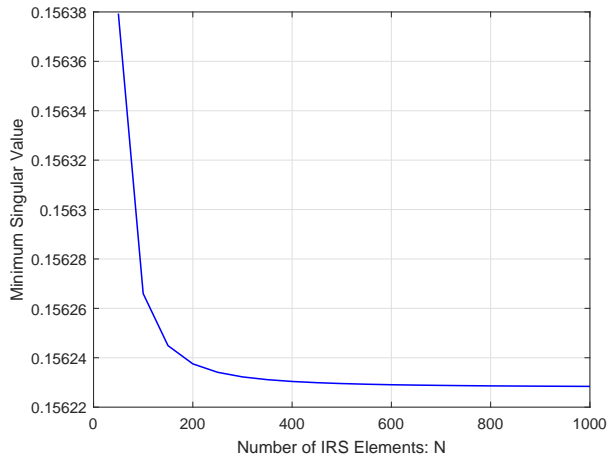}}}\subfigure[Validation under Covariance Model 2.]{\scalebox{0.58}{\includegraphics*{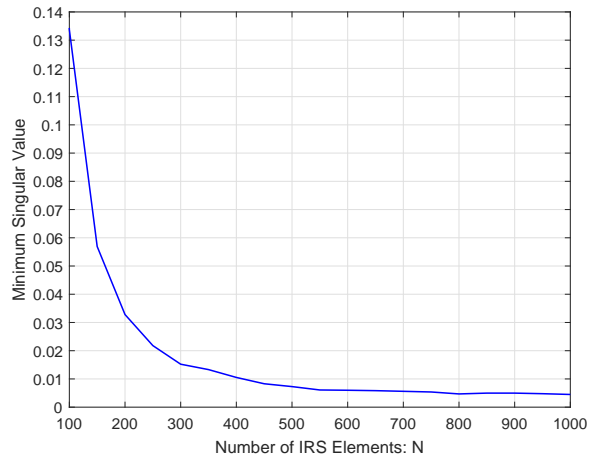}}}
	\end{center}
	\caption{Minimum singular value of $\mv{L}_1$.}\label{Dk}\vspace{-10pt}
\end{figure}

\begin{figure}
	\begin{center}
		\subfigure[$N=100$ under Covariance Model 1.]{\scalebox{0.5}{\includegraphics*{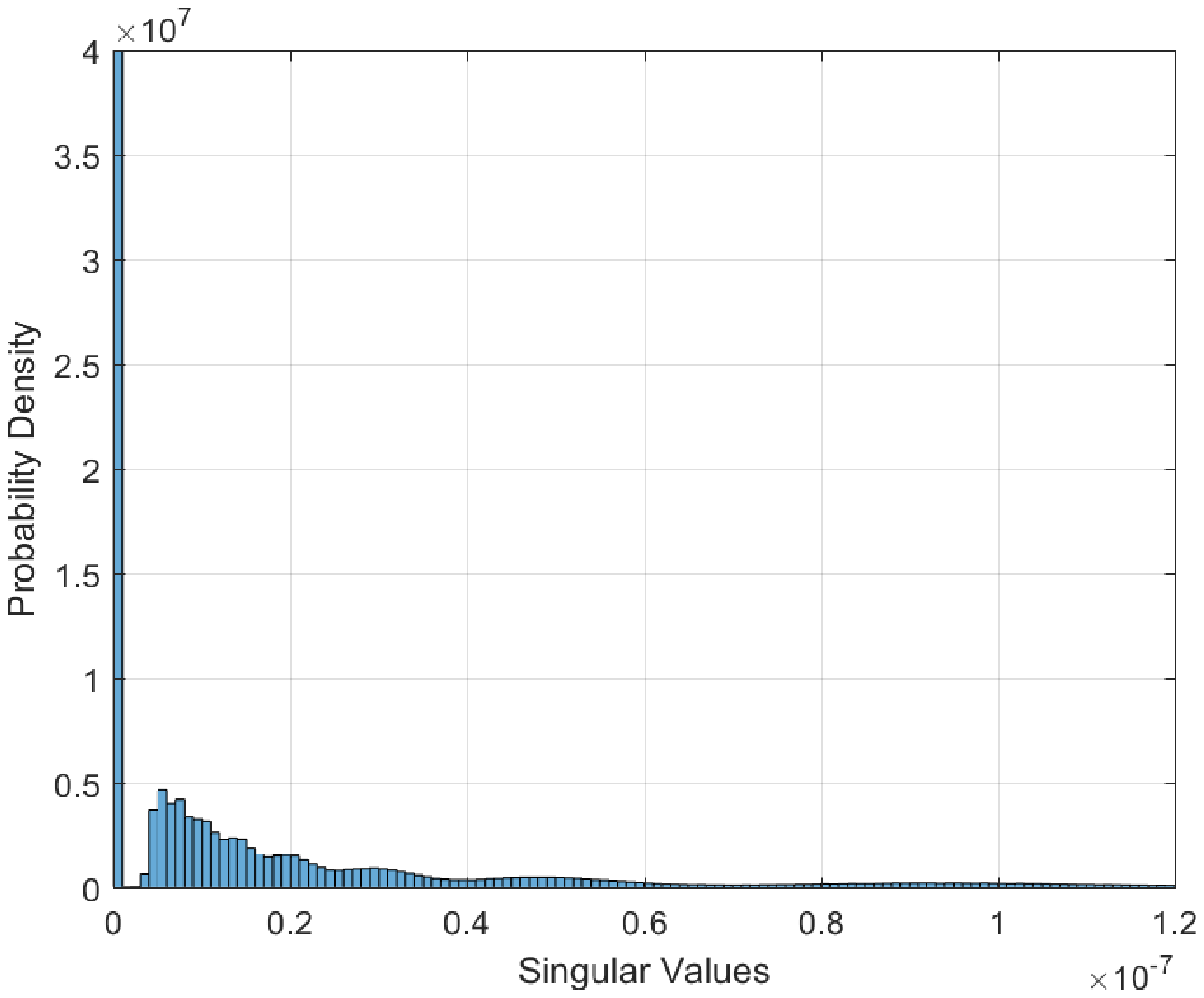}}}
		\subfigure[$N=200$ under Covariance Model 1.]{\scalebox{0.5}{\includegraphics*{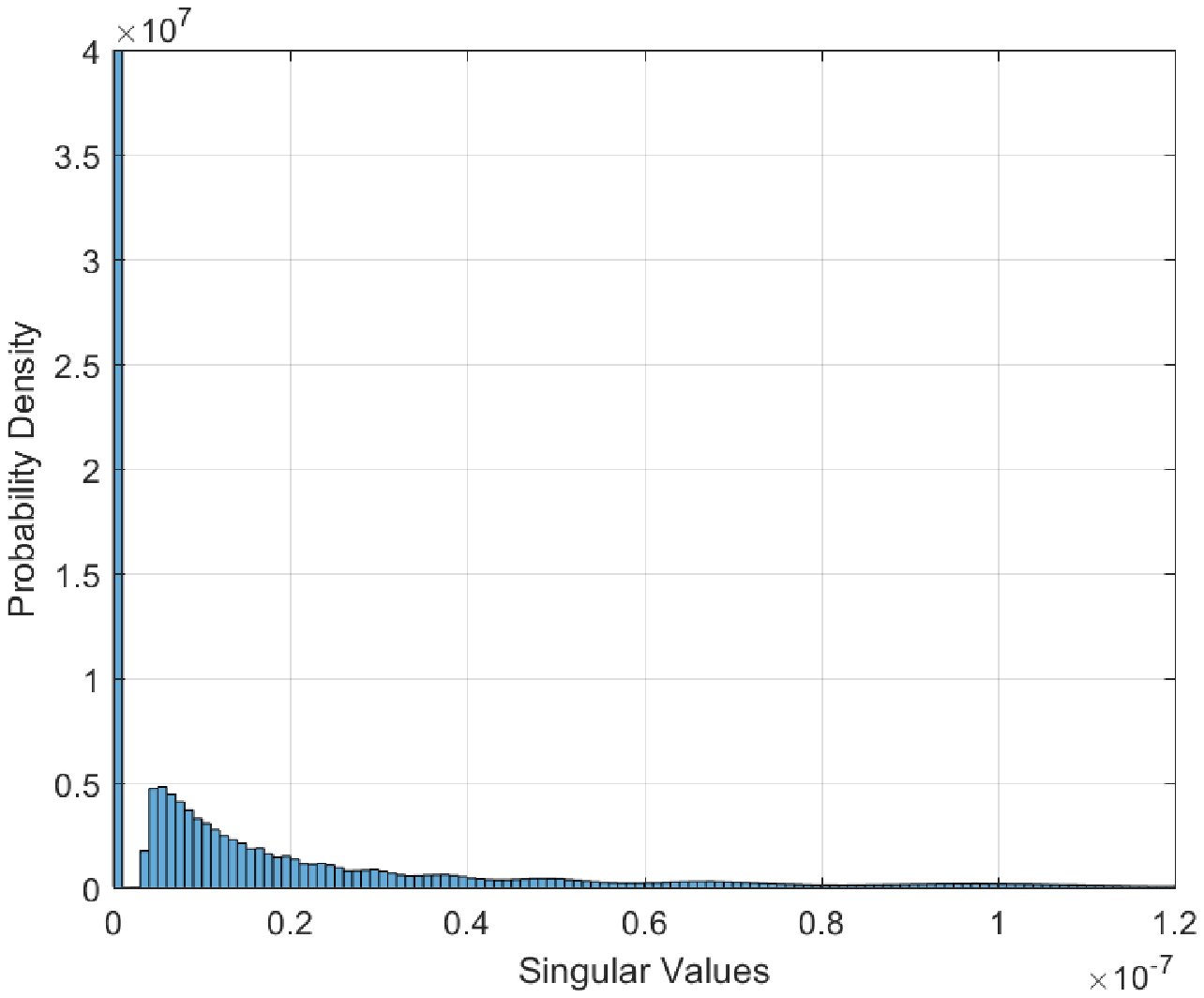}}}
		\subfigure[$N=100$ under Covariance Model 2.]{\scalebox{0.5}{\includegraphics*{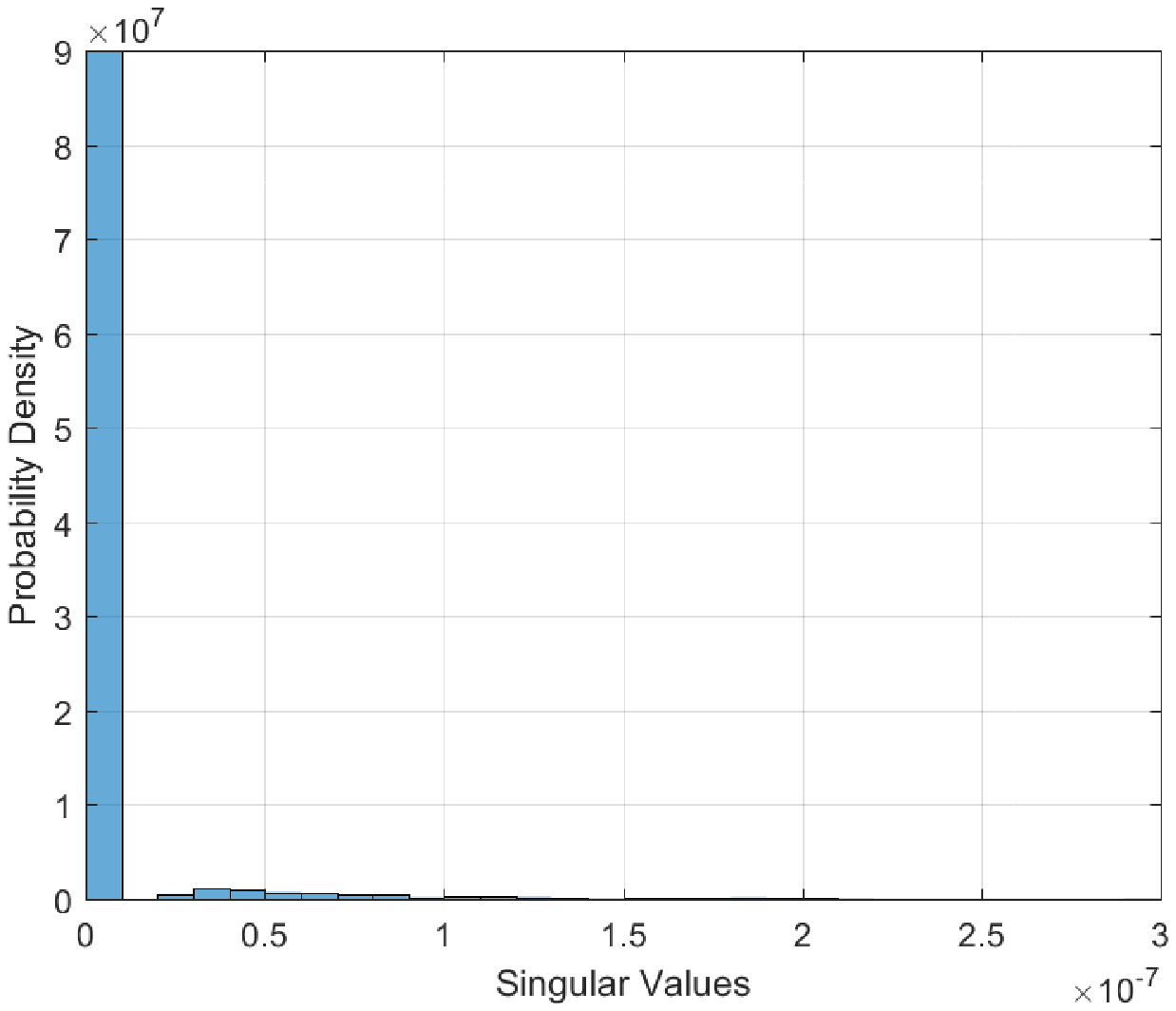}}}
		\subfigure[$N=200$ under Covariance Model 2.]{\scalebox{0.5}{\includegraphics*{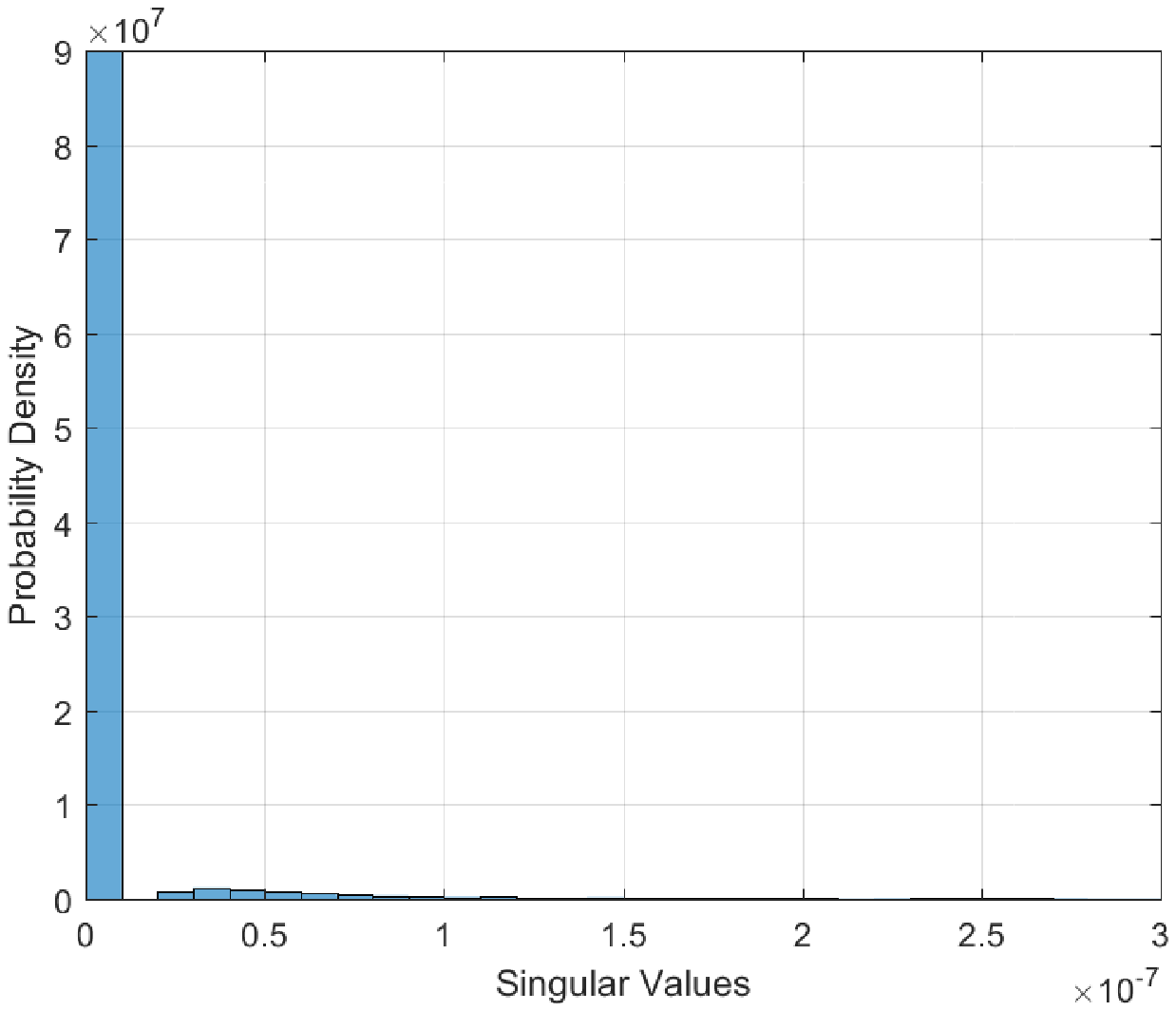}}}
	\end{center}
	\caption{Empirical distribution of the singular values of $\tilde{\mv{C}}_{1,2}$. }\label{eig-distri} \vspace{-20pt}
\end{figure}

Assumption \ref{assumption1} has been widely assumed in the massive MIMO works (see, e.g., \cite{hoydis2013massive}). For Assumptions \ref{assumption1.5} and \ref{assumption2}, it can be shown that when all the channel correlation matrices are identity matrices, i.e., $\mv{C}_k^{{\rm I}}=\mv{C}_k^{{\rm B}}=\mv{C}^{{\rm I}}=\mv{C}^{{\rm B}}=\mv{I}$, $\forall k$, these assumptions are true. Otherwise, it is hard to theoretically show the sufficient conditions for Assumptions \ref{assumption1.5} and \ref{assumption2} to be true. In the following, we verify these two assumptions numerically under the following two correlation models that are widely used in the literature:

\underline{\emph{Covariance Model 1}}: All the channel covariance matrices follow the exponential covariance matrix model\cite{Loyka2001capacity}, i.e., for a covariance matrix $\mv{C}$, the element at the $i$-th row and $j$-th column of $\mv{C}$ is expressed as $[\mv{C}]_{i,j}=c^{i-j}$ if $i\ge j$, and $[\mv{C}]_{i,j}=[\mv{C}]_{j,i}^*$ if $i<j$, with $|c|<1$.

\underline{\emph{Covariance Model 2}}: $\mv{C}^{\rm B}$ and $\mv{C}_k^{\rm B}$'s follow the exponential covariance matrix model\cite{Loyka2001capacity}, and $\mv{C}^{\rm I}$ and $\mv{C}_k^{\rm I}$'s follow the model shown in Proposition 1 of \cite{bjornson2020rayleigh}.

In this numerical example, we assume that $\phi_n=e^{j\theta_n}$, where $\theta_n$ is uniformly distributed in $[0 ~ 2\pi)$, $\forall n$. We first numerically verify Assumption \ref{assumption1.5}.  Fig. \ref{Dk} shows the minimum singular value of $\mv{L}_1$ under the above two channel covariance models. It is observed that as $N$ increases from 50 to 1000, the minimum singular value of $\mv{L}_1$ is lower bounded by a threshold which does not further decreases with $N$. We also have similar observations for $\mv{L}_k$, $k\ne1$, and for other kinds of correlation matrices construction, the details are omitted due to the space limit. Next, we numerically verify Assumption \ref{assumption2}. In this numerical example, we generate a large number of  $\tilde{\mv{R}}$ randomly according to its distribution. Fig. \ref{eig-distri} shows the empirical distribution of the singular values of $\tilde{\mv{C}}_{1,2}$ when  $N=100, 200$ and $q=N/M=10$ under the above two channel covariance models. It is observed that as $N$ and $M$ increase, the empirical distribution of the singular values approaches to a fixed distribution, and all the singular values fall into a fixed interval with probability 1. It implies that the spectral norm of $\tilde{\mv{C}}_{1,2}$ is indeed uniformly bounded almost surely in this numerical example. We also have similar observations for $\tilde{\mv{C}}_{i,j}$'s and $\hat{\mv{C}}_{i,j}$'s when $(i,j)\neq (1,2)$, the details  are omitted due to the space limit.

In the rest of this paper, we show the performance of the two-timescale protocol under Assumptions \ref{assumption1} to \ref{assumption2} and the assumption that the channel covariance matrices have already been estimated via the method proposed in \cite{liang2001downlink}. First, given $\phi_n$'s, we will show how to estimate $\mv{c}_k$'s at the beginning of each coherence block in Sections \ref{sec:channel estimation} and how to characterize the closed-form user achievable rates under the MRC receive beamforming strategy with imperfect CSI about $\mv{c}_k$'s in Section  \ref{sec:channel estimation}, respectively. Then, based on the user rate characterization as functions of $\phi_n$'s, we will design an efficient algorithm to optimize $\phi_n$'s to minimize the users' total transmit power over the whole covariance interval in Section \ref{sec:design phi}.

\section{Channel Estimation Design}\label{sec:channel estimation}
In this section, for given  $\phi_n$'s under our considered communication protocol, we  introduce MMSE channel estimators of the effective channels $\mv{c}_k$'s.

In the channel estimation stage, each user $k$ sends a pilot sequence with length $\tau=K$\cite{Hassibi03} to the BS, denoted by $\mv{a}_k=[a_{k,1},\cdots,a_{k,K}]^T$,  where $a_{k,i}$ with $|a_{k,i}|=1$ denotes the $i$-th pilot symbol of user $k$, $i=1,...,K$. Assuming that all the users transmit their pilots with identical power denoted by $p^{\rm t}$,  the normalized received signal at the BS during the channel training stage is
\begin{align}
	\mv{Y}/\sqrt{N}=\sqrt{p^{\rm t}}\left([\mv{c}_1,\cdots,\mv{c}_K]/\sqrt{N}\right)\left[\mv{a}_1,\dots,\mv{a}_K\right]^T+\mv{Z}/\sqrt{N}, \label{eq:rec}
\end{align}
where $\mv{Z}\sim \mathcal{CN}(\mv{0},K\sigma^2\mv{I})\in \mathbb{C}^{M\times K}$ denotes the AWGN. Here, we normalize the received signal by $\sqrt{N}$ to ease the analysis in the asymptotic regime \eqref{eqn:regime} later. Since the cross-covariance matrix of any two different normalized channels $\mv{c}_k/\sqrt{N}$ and $\mv{c}_j/\sqrt{N}$  is
\begin{align}
	&\mathbb{E}\left[\frac{\mv{c}_k\mv{c}_j^H}{N}\right]\nonumber\\
	&=\mathbb{E}\left[\frac{\mv{h}_k\mv{h}_j^H}{N}\right]+\mathbb{E}\left[\frac{\left(\mv{C}^{\rm B}\right)^{\frac{1}{2}}\tilde{\mv{R}}(\mv{C}^{\rm I})^{\frac{1}{2}}{\rm diag}(\mv{\phi})(\mv{C}_k^{\rm I})^{\frac{1}{2}}\tilde{\mv{t}}_k(\tilde{\mv{t}}_j)^H(\mv{C}_j^{\rm I})^{\frac{1}{2}}({\rm diag}(\mv{\phi}))^H(\mv{C}^{\rm I})^{\frac{1}{2}}(\tilde{\mv{R}})^H\left(\mv{C}^{\rm B}\right)^{\frac{1}{2}}}{N}\right]\nonumber\\
	&=\mathbb{E}\left[\frac{\left(\mv{C}^{\rm B}\right)^{\frac{1}{2}}\tilde{\mv{R}}(\mv{C}^{\rm I})^{\frac{1}{2}}{\rm diag}(\mv{\phi})(\mv{C}_k^{\rm I})^{\frac{1}{2}}\mathbb{E}\left[\tilde{\mv{t}}_k(\tilde{\mv{t}}_j)^H\right](\mv{C}_j^{\rm I})^{\frac{1}{2}}({\rm diag}(\mv{\phi}))^H(\mv{C}^{\rm I})^{\frac{1}{2}}(\tilde{\mv{R}})^H\left(\mv{C}^{\rm B}\right)^{\frac{1}{2}}}{N}\right]\nonumber \\ &=\mv{0}, ~~ \forall k\ne j,\label{eq:corre}
\end{align}
we follow the optimal pilot design such that the pilot sequences  of different users  are orthogonal with each other \cite{Hassibi03}, i.e., $(\mv{a}_k)^T\mv{a}_j=0$, $\forall k\neq j$. In this case,  from \eqref{eq:rec}, we have
\begin{align}
	\hat{\mv{y}}_k=\mv{Y}\mv{a}_k^\ast/\sqrt{N}=K\sqrt{p^{\rm t}}\mv{c}_k/\sqrt{N}+\hat{\mv{z}}_k, \label{eq:rec2}
\end{align}
where $\hat{\mv{z}}_k=\mv{Z}\mv{a}_k^\ast/\sqrt{N}\sim \mathcal{CN}(\mv{0},K\sigma^2\mv{I}/N)$. The MMSE channel estimator of $\mv{c}_k/\sqrt{N}$ is
\begin{align}
\hat{\mv{c}}_k=\mathbb{E}\left(\frac{\mv{c}_k}{\sqrt{N}}\bigg\vert\hat{\mv{y}}_k\right), ~~ k=1,\dots,K,	\label{eq:MMSE}
\end{align}
where $\mathbb{E}(x)$ denotes expectation of $x$. Define the error for the above MMSE estimation of $\mv{c}_k/\sqrt{N}$ as $\mv{\varepsilon}_k=\hat{\mv{c}}_k-\frac{\mv{c}_k}{\sqrt{N}}$, $\forall k$. Then, the MSE is
\begin{align}
\mv{F}_k=\mathbb{E}\left(\mv{\varepsilon}_k\mv{\varepsilon}_k^H\right),~~~\forall k. \label{eq:F_k_ini}
\end{align}
In general, it is hard to obtain the closed-form expressions of the MMSE channel estimators \eqref{eq:MMSE} and the corresponding estimation MSE \eqref{eq:F_k_ini} due to the complicated channel distribution arising from the deployment of the IRS. Nevertheless, in the rest of this section, we show that in our considered asymptotic regime \eqref{eqn:regime}, $\mv{c}_k/\sqrt{N}$'s tend to be Gaussian distributed such that they can be efficiently estimated based on the classic estimation techniques for Rayleigh fading channels.

\subsection{Channel Distribution and Channel Properties in Asymptotic Regime}\label{sec:gaussian}
We first study the Gaussian approximation for the distribution of the normalized effective channels
\begin{align}\label{eqn:c}
\mv{c}=\left[\mv{c}_1^T/\sqrt{N},\dots,\mv{c}_K^T/\sqrt{N}\right]^T,
\end{align}in \eqref{eq:rec} under the asymptotic regime (\ref{eqn:regime}). For convenience, define the reflecting channel related term within $\mv{c}_k$ shown in \eqref{eq: c} as
\begin{align}
	\mv{q}_k=[q_{k,1},\ldots,q_{k,M}]^T=\frac{(\mv{C}^{\rm B})^{-\frac{1}{2}}}{\sqrt{N}}\sum_{n=1}^{N}\phi_{n}\mv{g}_{k,n}=\frac{1}{\sqrt{N}}\tilde{\mv{R}}(\mv{C}^{\rm I})^{\frac{1}{2}}{\rm diag}(\mv{\phi})(\mv{C}_k^{\rm I})^{\frac{1}{2}}\tilde{\mv{t}}_k, ~~\forall k. \label{eq: g}
\end{align}The following lemma shows that the vectors $\mv{q}_1,\ldots,\mv{q}_K$ are jointly Gaussian vectors under Assumption \ref{assumption1.5} when $N$ goes to infinity.

\begin{lemma}\label{lemma1}
Denote $\mv{q}=\left[\mv{q}_1^T,\dots,\mv{q}_K^T\right]^T$, where $\mv{q}_k$'s are given in (\ref{eq: g}). Under Assumption \ref{assumption1.5}, $\mv{q}$ is a Gaussian random vector when $N$ goes to infinity.
\end{lemma}

\begin{IEEEproof}
Please refer to Appendix \ref{appendix4}.
\end{IEEEproof}

Define
\begin{align}
	\bar{\mv{q}}={\rm diag}\left(\left(\mv{C}^{\rm B}\right)^{\frac{1}{2}},\dots,\left(\mv{C}^{\rm B}\right)^{\frac{1}{2}}\right)\mv{q}.
\end{align}Since $\mv{q}$ is a Gaussian random vector according to Lemma \ref{lemma1}, $\bar{\mv{q}}$ is also a Gaussian vector. Further, $\mv{h}=\left[\mv{h}_1^T,\dots,\mv{h}_K^T\right]^T$ is a Gaussian random vector which is independent of $\bar{\mv{q}}$. As a result, according to (\ref{eq: c}) and (\ref{eqn:c}), $\mv{c}=\mv{h}/\sqrt{N}+\bar{\mv{q}}$ is a  Gaussian vector. In this case, we have  $\mv{c}_k/\sqrt{N}\in \mathcal{CN}(\mv{0},\mv{V}_k/N)$ with $\mv{V}_k$ given in \eqref{eq: V_k}, $\forall k$.
Last, since the jointly Gaussian vectors $\mv{c}_1/\sqrt{N},\dots,\mv{c}_K/\sqrt{N}$ are uncorrelated as shown in \eqref{eq:corre}, they are independent of each other. To summarize, Lemma \ref{lemma1} implies the following corollary.
\begin{corollary}
Under Assumption \ref{assumption1.5}, $\mv{c}_1/\sqrt{N},\dots,\mv{c}_K/\sqrt{N}$, where $\mv{c}_k$'s are given in (\ref{eq: c}), are independent and jointly Gaussian random vectors when $N$ goes to infinity.
\end{corollary}

\begin{figure}[ht]
	\centering
	\subfigure[Distribution of $\mv{v}_1$ under Covariance Model 1.]
	{
	 \label{apdffit}
	 \includegraphics[width=0.47\columnwidth]{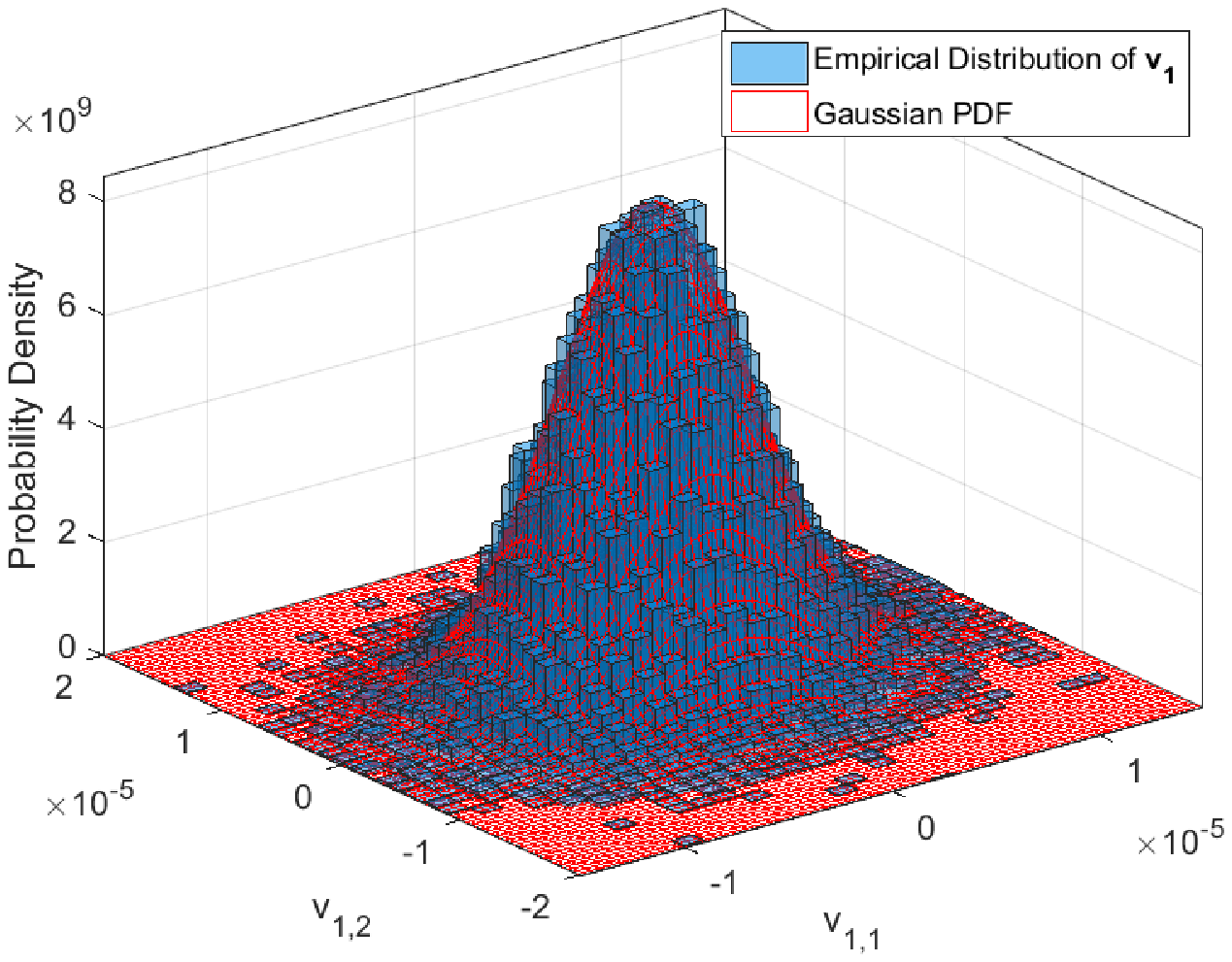}
    }
	\subfigure[Distribution of $\mv{v}_2$ under Covariance Model 1.]
	{
     \label{bpdffit}
	 \includegraphics[width=0.47\columnwidth]{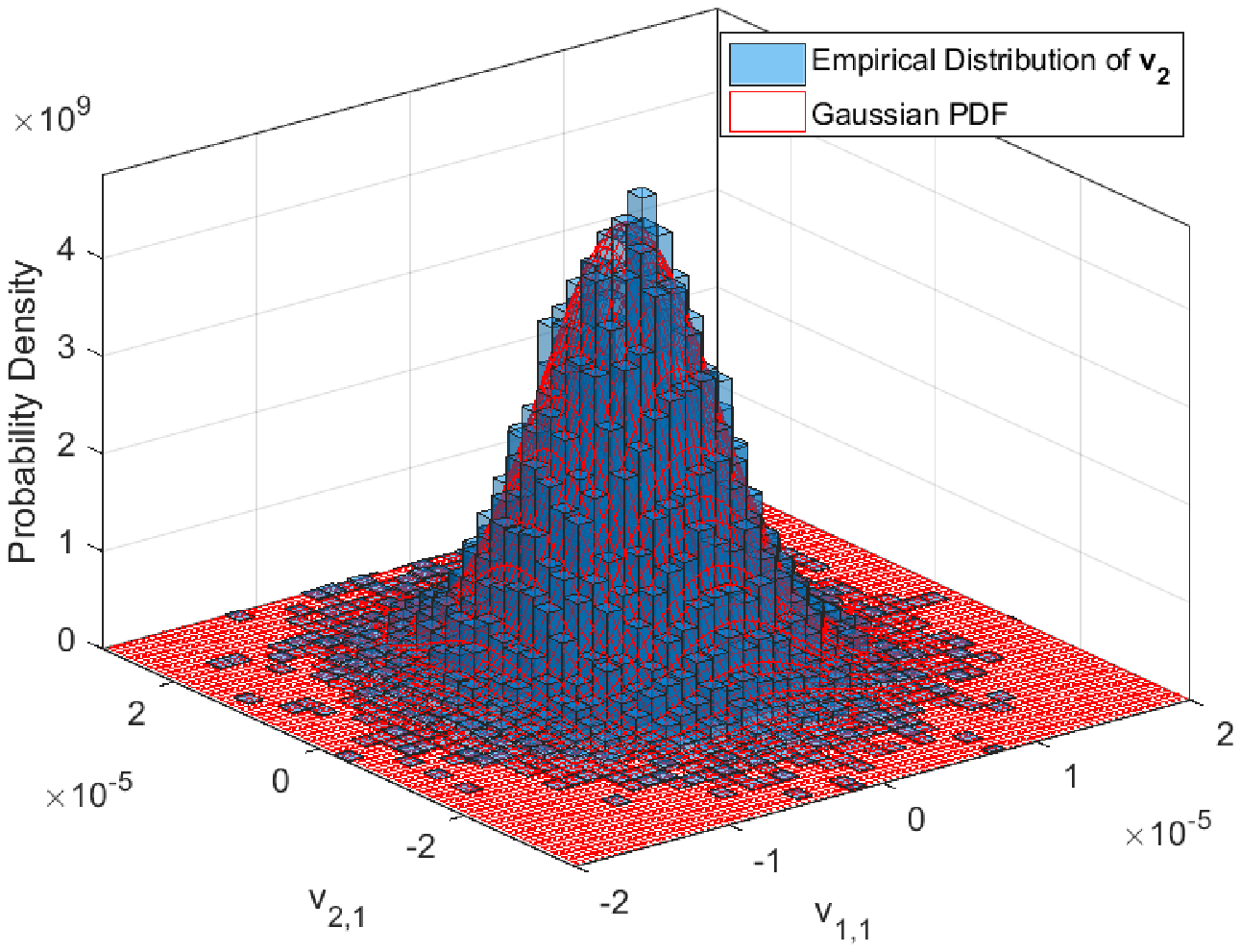}
    }
   \subfigure[Distribution of $\mv{v}_1$ under Covariance Model 2.]
   {
	\label{cpdffit}
	\includegraphics[width=0.47\columnwidth]{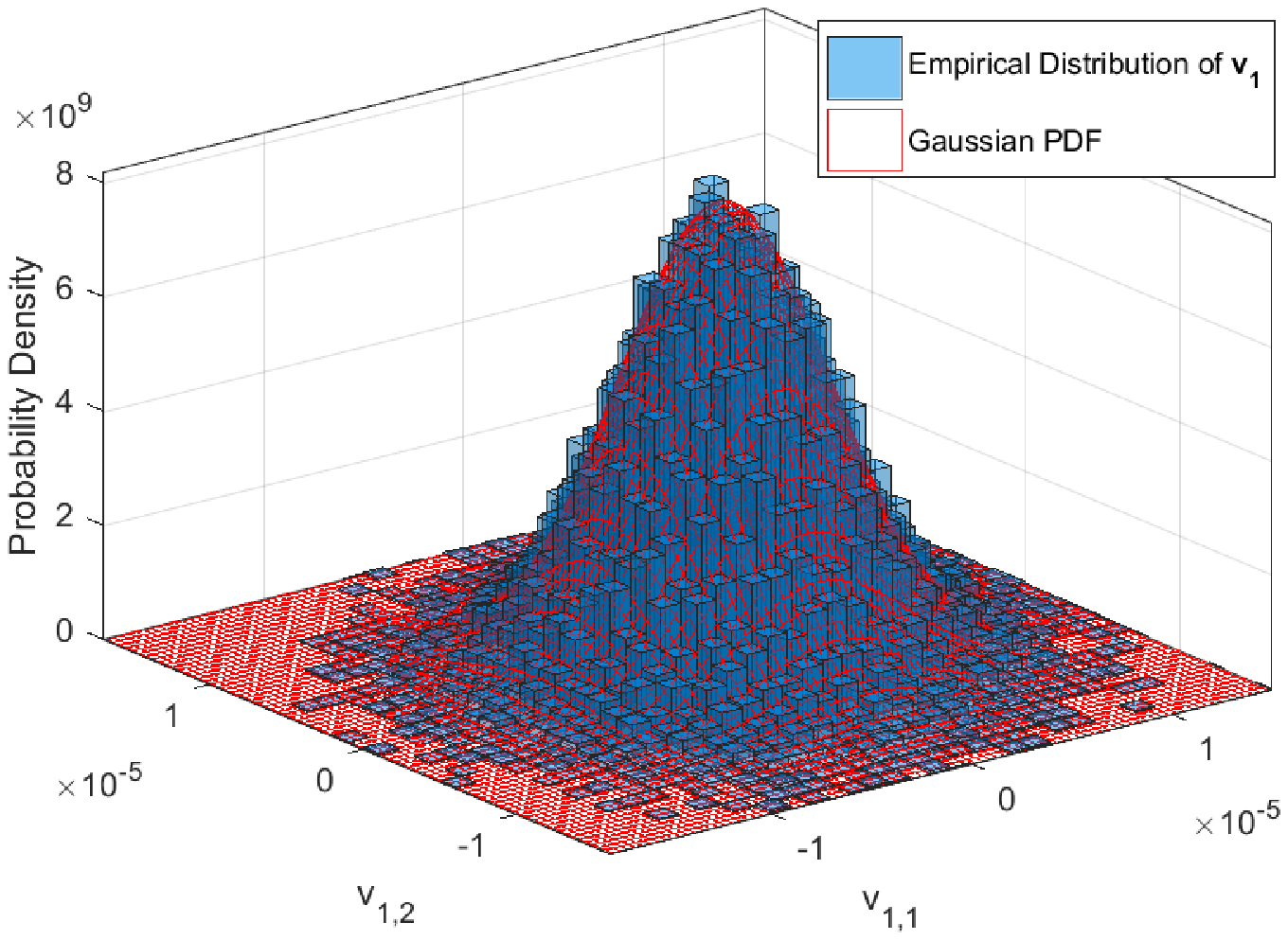}
   }
   \subfigure[Distribution of $\mv{v}_2$ under Covariance Model 2.]
   {
	\label{dpdffit}
	\includegraphics[width=0.47\columnwidth]{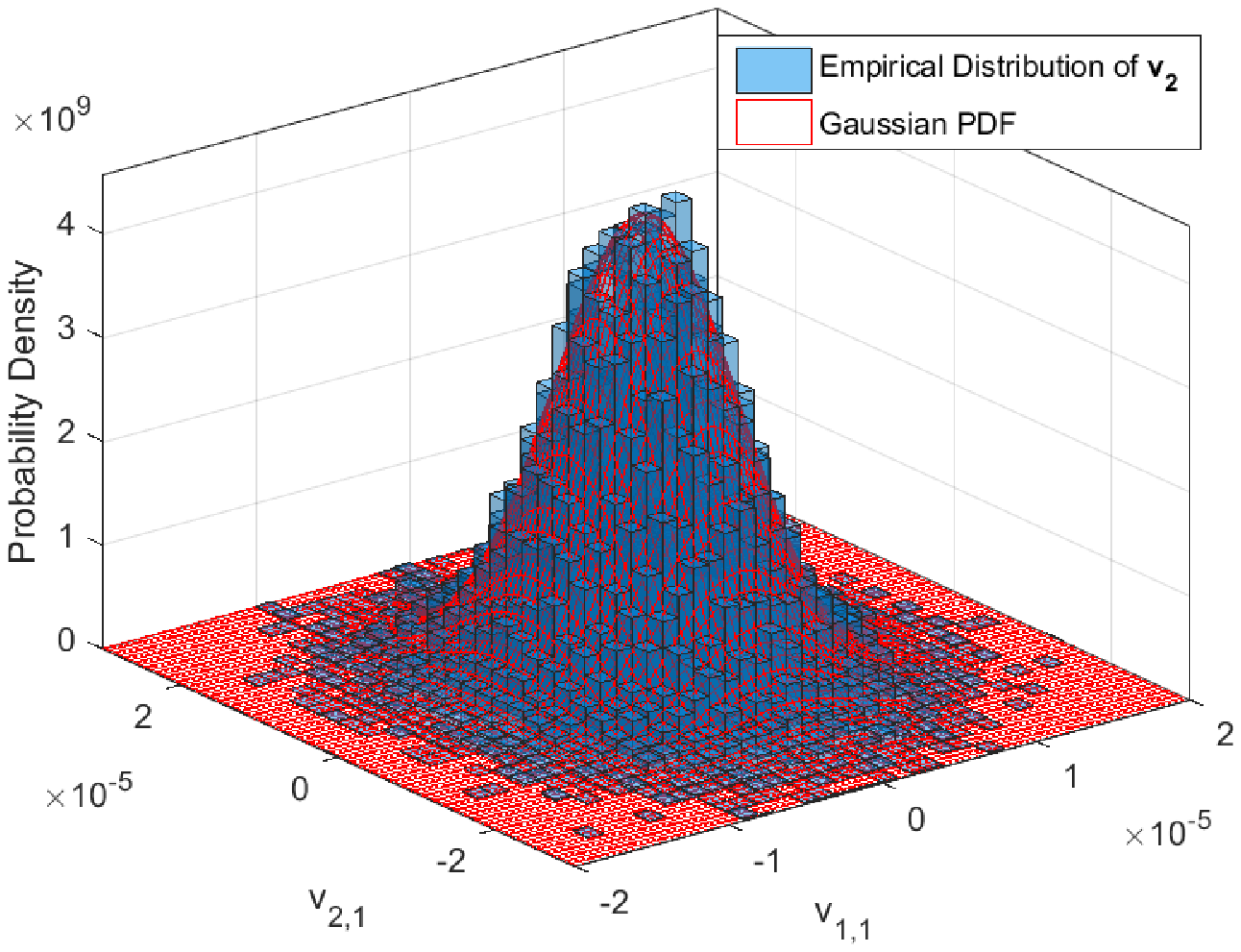}
    }
	\caption{Verification of the Gaussian approximation for the distribution of $\mv{v}_1$ and $\mv{v}_2$.}\label{pdffit}
\end{figure}

In the following, we provide a numerical example to verify the tightness of the Gaussian distribution approximation  of $\left[\mv{c}_1^T/\sqrt{N},\mv{c}_2^T/\sqrt{N}\right]^T$. Define $v_{1,1}$ and $v_{1,2}$ as the real parts of the first and the second elements in $\mv{c}_1/\sqrt{N}$, and $v_{2,1}$ as the real part of first element in $\mv{c}_2/\sqrt{N}$, respectively. Under the above setup, in Figs. \ref{apdffit} - \ref{dpdffit}, we plot the empirical distribution of $\mv{v}_1=[v_{1,1},v_{1,2}]^T$ and $\mv{v}_2=[v_{1,1},v_{2,1}]^T$ when $N=300$ and $q=10$, under the two covariance models as detailed in Section \ref{sec:protocol design}, respectively. Moreover, we plot the probability density functions (PDFs) of Gaussian random vectors with zero mean and covariance matrices  $\mv{\Sigma}_1$ and  $\mv{\Sigma}_2$ to evaluate the empirical distribution of $\mv{v}_1$ and $\mv{v}_2$, respectively. It is observed from Fig. \ref{pdffit} that the empirical distribution of $\mv{v}_1$ and $\mv{v}_2$  does matches that of the Gaussian random vectors very well when $N=300$ and $q=10$. Similar results are observed in many other numerical results, which are omitted here due to the space limit.

We next study the channel hardening and favorable propagation properties for the user effective channels in our considered IRS-assisted massive MIMO systems. We have the following theorem.
\begin{theorem}\label{theorem0}
	In the asymptotic regime where $N$ and $M$ go to infinity with a fixed ratio $N/M=q$, the following properties are true for the normalized effective channels $\mv{c}_k/\sqrt{N}$'s under Assumptions \ref{assumption1} to \ref{assumption2}:
	\begin{align}
		&\frac{\mv{c}_k^H\mv{c}_k}{MN}-\frac{\beta^{\rm IU}_k\beta^{\rm BI}\mv{\phi}^H\bar{\mv{C}}_k\mv{\phi}}{N}\xrightarrow{\text{a.s.}}0,~\forall k,\label{eq: chan-hard}\\
		&\frac{\mv{c}_k^H\mv{c}_j}{MN}\xrightarrow{\text{a.s.}}0, ~~\forall k\ne j, \label{eq: chan-fav}
	\end{align}
where  ``$a\xrightarrow{\text{a.s.}}b$'' denotes ``almost sure convergence of $a$ to $b$''.
\end{theorem}
\begin{IEEEproof}
	Please refer to Appendix \ref{appendix2}.
\end{IEEEproof}

In the conventional massive MIMO systems without IRS, the so-called ``channel hardening'' and ``favorable propagation'' properties hold, which state that as $M$ goes to infinity, the impact of the channel randomness on the communications is negligible compared with its channel power, and the user channels are orthogonal with each other. Interestingly,  the channel hardening (see  \eqref{eq: chan-hard}) and favorable propagation (see \eqref{eq: chan-fav}) properties still hold in the IRS-assisted massive MIMO system, if $M$ and $N$ go to infinity with a fixed ratio $N/M=q$.  Moreover, \eqref{eq: chan-hard} also explains the reason that we normalize the effective channels $\mv{c}_k$'s (and the received signals) by $\sqrt{N}$. Specifically, as shown in \eqref{eq: chan-hard}, the power of the normalized channel $\mv{c}_k/\sqrt{N}$ is in the order of $M$, and the  power of each normalized channel element in $\mv{c}_k/\sqrt{N}$ converges to a fixed value, $k=1,\dots,K$, making the afterwards signal processing doable even through the number of IRS elements $N$ goes to infinity.

\subsection{MMSE Channel Estimators and Corresponding Estimation MSE in the Asymptotic Regime}
After quantifying the distribution of $\mv{c}_k/\sqrt{N}$'s in our considered asymptotic regime, in this subsection, we study the MMSE estimators of $\mv{c}_k/\sqrt{N}$'s in \eqref{eq:MMSE} and  the estimation MSE in \eqref{eq:F_k_ini} under the regime. Since $\mv{c}_k/\sqrt{N}$'s are independent Gaussian random vectors shown in Section \ref{sec:gaussian}, the MMSE estimator is equivalent to a simple LMMSE estimator.  In this case, under the considered regime \eqref{eqn:regime}, the MMSE channel estimator of $\mv{c}_k/\sqrt{N}$ in \eqref{eq:MMSE} becomes
\begin{align}
\hat{\mv{c}}_k=\mv{V}_k\left(\mv{V}_k+\frac{\sigma^2}{Kp^t}\mv{I}\right)^{-1}\left(\mv{c}_k+\frac{1}{K\sqrt{p^t}}\tilde{\mv{z}}_k\right)\bigg/\sqrt{N},~ \forall k, \label{eq:estimated channel}
\end{align}
where $\tilde{\mv{z}}_k=\mv{Z}\mv{a}_k^\ast\in\mathcal{CN}\left(\mv{0},K\sigma^2\mv{I}\right)$, and $\mv{V}_k$ is defined in \eqref{eq: V_k}. In addition, the MSE in \eqref{eq:F_k_ini} becomes
\begin{align}
\mv{F}_k=\mathbb{E}\left(\mv{\varepsilon}_k\mv{\varepsilon}_k^H\right)=\left(\mv{V}_k^{-1}+\frac{Kp^t}{\sigma^2}\mv{I}\right)^{-1}\bigg/\sqrt{N},~~~\forall k. \label{eq:F_k}
\end{align}

In the following, we study the channel estimation error. In \eqref{eq:F_k}, denote the eigenvalue decomposition of  $\mv{V}_k$  as
\begin{align}
	\mv{V}_k=\mv{B}_k\mv{\Sigma}_k\mv{B}_k^H, \label{eq:evd_v}
\end{align}
where  $\mv{B}_k\in\mathbb{C}^{M\times M}$ is a unitary matrix, and $\mv{\Sigma}_k={\rm diag}(\lambda_{k,1},\dots,\lambda_{k,M})$ is a diagonal matrix whose diagonal elements are $M$ eigenvalues. Substituting \eqref{eq:evd_v} into \eqref{eq:F_k}, the power of $\mv{\varepsilon}_k$ is
\begin{align}
	\mathbb{E}\left(\mv{\varepsilon}_k^H\mv{\varepsilon}_k\right)={\rm tr}\left(\mv{F}_k\right)&=\frac{1}{N}{\rm tr}\left(\mv{B}_k^H{\rm diag}\left(\frac{\lambda_{k,1}\sigma^2}{\sigma^2+p^tK\lambda_{k,1}},\dots,\frac{\lambda_{k,M}\sigma^2}{\sigma^2+p^tK\lambda_{k,M}}\right)\mv{B}_k\right)\label{eq:Fk_de}\\
	&=\frac{1}{N}\sum_{i=1}^M\frac{\lambda_{k,i}\sigma^2}{\sigma^2+p^tK\lambda_{k,i}}=\frac{\sigma^2}{Np^tK}\sum_{i=1}^M\frac{p^tK\lambda_{k,i}}{\sigma^2+p^tK\lambda_{k,i}}\le \frac{M\sigma^2}{Np^tK}.
\end{align}
In this case,
\begin{align}
	0\le \lim_{N/M=q, N\rightarrow \infty} \frac{{\rm tr}\left(\mv{F}_k\right)}{M}\le\lim_{N/M=q, N\rightarrow \infty}\frac{\sigma^2}{Np^tK}=0.
\end{align}
Thus,
\begin{align}
	\lim_{N/M=q, N\rightarrow \infty} \frac{{\rm tr}\left(\mv{F}_k\right)}{M}=0. \label{eq:Fk2}
\end{align}
Note that \eqref{eq:Fk2} indicates that in the IRS-assisted massive MIMO system, the channel estimation error is negligible compared to the normalized channel power order $M$. Note that this is not true in the conventional massive MIMO system without IRS\cite{ngo2013energy}. The reason is as follows. Based on \eqref{eq:rec2}, since the elements in $\mv{c}_k/\sqrt{N}$ and $\hat{\mv{z}}_k$ are independent, the channel of each antenna $m$ can be estimated via the following signal
\begin{align}
	\hat{\mv{y}}_k(m)=K\sqrt{p^{\rm t}}\mv{c}_k(m)/\sqrt{N}+\hat{\mv{z}}_k(m),~~ m=1,\dots,M, \label{eq:rec3}
\end{align}
where $\hat{\mv{z}}_k(m)\sim \mathcal{CN}(0,K\sigma^2/\sqrt{N})$. It is observed that as $N$ goes to infinity, the power of $\mv{c}_k(m)/\sqrt{N}$ is finite as shown in \eqref{eq: chan-hard} since it is the superposition of $N$ reflecting channels, but the power of $\hat{\mv{z}}_k(m)$ is zero. As a result, with the MMSE channel estimator \eqref{eq:estimated channel}, the MSE for estimation $\mv{c}_k(m)/\sqrt{N}$ is negligible compared to the power of $\mv{c}_k(m)/\sqrt{N}$. Thus,
\begin{align}
	\hat{\mv{c}}_k(m)=\frac{\mv{c}_k(m)}{\sqrt{N}}, ~~m=1,\dots,M.
\end{align}
As a result, it is expected that the overall MSE for estimating $\mv{c}_k/\sqrt{N}$ is negligible compared to the power of $\mv{c}_k/\sqrt{N}$, $k=1,\dots,K$, as shown in \eqref{eq:Fk2}.

A straightforward result following from \eqref{eq:Fk2}, and the channel hardening in \eqref{eq: chan-hard} and favorable propagation property in  \eqref{eq: chan-fav} is that, in the asymptotic regime where $N$ and $M$ go to infinity with a fixed ratio $N/M=q$, we have
\begin{align}
	&\frac{\hat{\mv{c}}_k^H\hat{\mv{c}}_k}{M}-\frac{\beta^{\rm IU}_k\beta^{\rm BI}\mv{\phi}^H\bar{\mv{C}}_k\mv{\phi}}{N}\xrightarrow{\text{a.s.}}0,~\forall k,	\label{eq:phi21}\\
	&\frac{\hat{\mv{c}}_k^H\hat{\mv{c}}_j}{M}\xrightarrow{\text{a.s.}}0, ~~\forall k\ne j.\label{eq:phi1}
\end{align}

\section{Achievable User Rate with Imperfect CSI}\label{sec:Date transmission}
In this section, we characterize the user achievable rate under the two-timescale protocol  in our considered asymptotic regime \eqref{eqn:regime}.

With the MMSE channel estimator $\hat{\mv{c}}_k$'s shown in \eqref{eq:MMSE} given any $M$ and $N$, the normalized beamformed signal  at the BS shown in \eqref{eq: y} can be expressed as
\begin{align}
\mv{\tilde{y}}_k/\sqrt{N}
=\mv{w}_k^H\hat{\mv{c}}_k\sqrt{p_k}s_k+\sum_{j\ne k}^K\mv{w}_k^H\hat{\mv{c}}_j\sqrt{p_j}s_j-\sum_{j=1}^{K}\mv{w}_k^H\mv{\varepsilon}_j\sqrt{p_j}s_j+\mv{w}_k^H\mv{z}/\sqrt{N},~~\forall k. \label{eq:y1}
\end{align}
Due to the channel hardening in \eqref{eq: chan-hard} and favorable propagation property in  \eqref{eq: chan-fav}, it can be expected that the MRC receiver is quite good. Thus, the BS applies MRC receivers based on the estimated channels, i.e., $\mv{w}_k=\hat{\mv{c}}_k$, $\forall k$. In this case, \eqref{eq:y1} becomes
\begin{align}
\mv{\tilde{y}}_k/\sqrt{N}
=\hat{\mv{c}}_k^H\hat{\mv{c}}_k\sqrt{p_k}s_k+\sum_{j\ne k}^K\hat{\mv{c}}_k^H\hat{\mv{c}}_j\sqrt{p_j}s_j-\sum_{j=1}^{K}\hat{\mv{c}}_k^H\mv{\varepsilon}_j\sqrt{p_j}s_j+\hat{\mv{c}}_k^H\mv{z}/\sqrt{N},~~\forall k.
\end{align}
Under MMSE channel estimation, the estimated channels $\hat{\mv{c}}_k$'s are independent of the estimation errors $\mv{\varepsilon}_k$'s. In this case, according to\cite{Hassibi03,ngo2013energy}, the achievable rate of user $k$ is
\begin{align}
R_k(\mv{\phi})=\frac{T-K}{T}\log_2(1+\gamma_k(\mv{\phi})), ~~\forall k,
\label{eq:Sys-1.6}
\end{align}
where $T$ is the length of a coherent block, $(T-K)/T$ denotes the fraction of time for data transmission, and $\gamma_k(\mv{\phi})$ is given by
\begin{align}
&\gamma_k(\mv{\phi})=\frac{p_k\left|\hat{\mv{c}}_k^H\hat{\mv{c}}_k\right|^2}{\sum_{j\ne k}^Kp_j\left|\hat{\mv{c}}_k^H\hat{\mv{c}}_j\right|^2+\sum_{j=1}^{K}p_j\hat{\mv{c}}_k^H\mv{F}_j\hat{\mv{c}}_k+\sigma^2\hat{\mv{c}}_k^H\hat{\mv{c}}_k/N},~~\forall k, \label{eq:Sys-A1.7}
\end{align}
with $\mv{F}_j$'s shown in \eqref{eq:F_k_ini}.

In the following, we characterize the closed-form expression of the user rates in \eqref{eq:Sys-A1.7} under the regime where $N$ and $M$ go to infinity with $N/M=q$. Within this regime, the closed-from expressions of $\hat{\mv{c}}_k$'s and $\mv{F}_k$'s are shown in \eqref{eq:estimated channel} and \eqref{eq:F_k}, respectively. We have the following theorem.
\begin{theorem}\label{theorem1}
	Assume that the transmit power of user $k$ is $p_k=\frac{E_k}{MN}$, where $E_k$ is fixed, $\forall k$. Under the asymptotic regime where $N$ and $M$ go to infinity with a fixed ratio $N/M=q$ and Assumptions \ref{assumption1} to \ref{assumption2}, we have
	\begin{align}
	R_k(\mv{\phi})-\bar{R}_k(\mv{\phi})\xrightarrow{\text{a.s.}}0,~~ \forall k,
	\end{align}where $R_k(\mv{\phi})$ is the rate of user $k$ given in \eqref{eq:Sys-1.6} achieved by the MRC beamforming $\mv{w}_k=\hat{\mv{c}}_k$, and
    \begin{align}
    \bar{R}_k(\mv{\phi})= \frac{T-K}{T}\log_2\left(1+\frac{E_k\beta^{\rm IU}_k\beta^{\rm BI}\mv{\phi}^H\bar{\mv{C}}_k\mv{\phi}}{N\sigma^2}\right), ~~\forall k,
    \label{eq:Sys-1.7}
    \end{align}
with $\bar{\mv{C}}_k$ being shown in \eqref{eq: C_k}.
\end{theorem}

\begin{IEEEproof}
Please refer to Appendix \ref{appendix3}.
\end{IEEEproof}

\begin{figure}[t]
	\begin{center}
		\subfigure[Covariance Model 1.]{\scalebox{0.58}{\includegraphics*{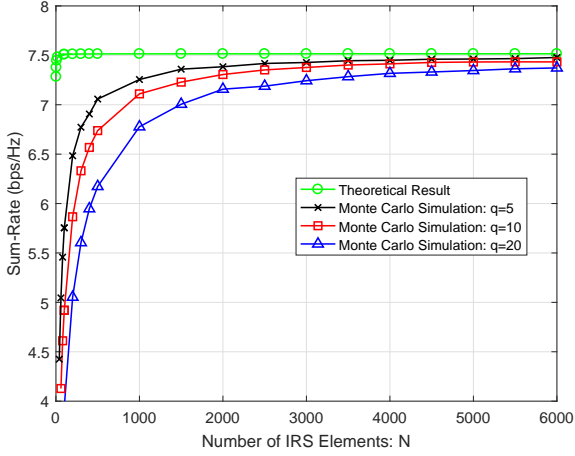}}}\subfigure[Covariance Model 2.]{\scalebox{0.58}{\includegraphics*{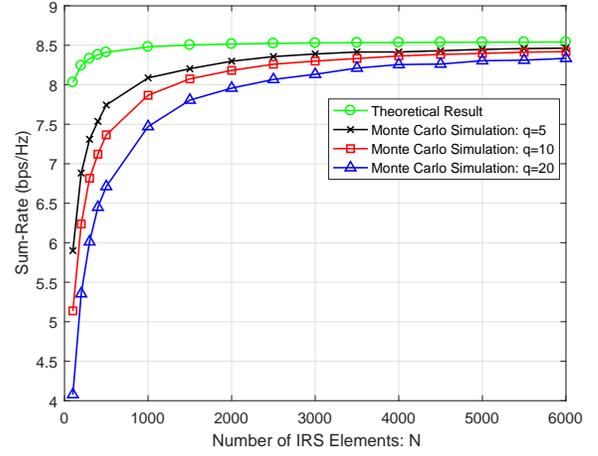}}}
	\end{center}
	\caption{Validation of Theorem \ref{theorem1} with different values of $q$ and channel covariance models.}\label{vali-theom1}\vspace{-20pt}
\end{figure}

We verify Theorem \ref{theorem1} through numerical results. Specifically, we assume that the number of users is $K=4$, and the ratio between the numbers of IRS reflection elements and BS antennas is $q=N/M=5, 10, 20$. The IRS reflection coefficients are set to be $\phi_n=1$, $\forall n$.  The transmit power is $p_k=\frac{E_k}{MN}$, where $E_k=0$ dbm  for all users. The power spectrum density of the AWGN is $-170$ dBm/Hz, and the bandwidth is assumed to be $1$ MHz. The channel fading block length is $T=1000$. We assume that the direct links between BS and users are totally blocked, i.e., $\mv{h}_k=\mv{0}$, $k=1,\dots, K$. Moreover, we apply the  two channel covariance models detailed in Section \ref{sec:protocol design} to verify Theorem \ref{theorem1}. The path loss of $t_{k,n}$'s and $\mv{r}_n$'s is modeled as  $\beta_k^{{\rm IU}}=\beta_0(d_k^{{\rm IU}}/d_0)^{-\alpha_1}$ and $\beta^{{\rm BI}}=\beta_0(d^{{\rm BI}}/d_0)^{-\alpha_2}$, respectively, where $d_0=1$ meter (m) denotes the reference distance, $\beta_0=-20$ dB denotes the path loss at the reference distance,  $d_k^{{\rm IU}}$ and $d^{{\rm BI}}$ denote the distance between  the IRS and user $k$, as well as the distance between the BS and the IRS,  $\alpha_1$ and $\alpha_2$ denote the path loss factors for  $t_{k,n}$'s and $\mv{r}_n$'s, respectively. We set  $\alpha_1=2.1$ and $\alpha_2=2.2$ in the numerical examples. Moreover, the distance between the BS and IRS is set to be $d^{\rm BI}=100$ m, and all the users are assumed to be located in a circular regime with radius $5$ m, whose center is $10$ m away from the IRS and $105$ m away from the BS.

Based on the above setup,  in Fig. \ref{vali-theom1} we plot the sum-rate predicated by \eqref{eq:Sys-1.7} and the sum-rate obtained by Monte Carlo simulation with different values of $q=N/M$. In the Monte Carlo simulations, we use LMMSE estimator to approximate MMSE estimator for general $M$ and $N$. It is observed from Fig. \ref{vali-theom1} that when $M$ and $N$ increase, the theoretical user rate predication in \eqref{eq:Sys-1.7} becomes more accurate under both of the two channel covariance models. Moreover, it is observed that with different values of $q$, the Monte Carlo simulations converge to the same rate, since \eqref{eq:Sys-1.7} implies that user rate is independent of $q$. However, with a smaller value of $q$ (or equivalently, a larger value of $M$), the user rate given any $N$ is closer to the asymptotic rate shown in \eqref{eq:Sys-1.7}. Fig. \ref{vali-theom1} indicates that with the aid of IRS, a non-zero rate is achieved when the user transmit power is in the order of $1/MN$. Note that in the conventional massive MIMO systems without IRS, a non-zero user rate is achieved when the user transmit power is in the order of $1/M$\cite{ngo2013energy}.

\section{IRS Reflection Coefficients Design}\label{sec:design phi}
In Sections \ref{sec:channel estimation} and \ref{sec:Date transmission}, we have introduced how to estimate the channels and characterize the user achievable rates given any IRS reflection coefficients in each channel coherent block, under our interested protocol introduced in Section \ref{sec:protocol design}. In this section, we introduce how to design the IRS reflection coefficients based on the channel covariance matrices to improve the user rates shown in \eqref{eq:Sys-1.7}. Due to the space limitation, in the following, we merely consider the problem to minimize the sum transmit power of all the users subject to the individual rate constraints, while the similar philosophy can be also applied to the other IRS reflection coefficients optimization problems, e.g., weighted sum-rate maximization problem, minimum rate maximization problem, etc.
Specifically, the sum-power minimization problem considered in this paper is formulated as
\begin{alignat}{3}
({\rm P1}):\quad &\min_{p_1,\dots,p_K,\mv{\phi}} &\quad &\sum\limits_{k=1}^{K} p_k \nonumber\\
&~~~\text{s.t.}                          &      & \bar{R}_k(\mv{\phi})\geq R_k^{\rm min}, ~~k=1,\dots, K,\label{eq:cons1}\\
&                                     &      & \phi_n^{\ast}\phi_n\le 1, ~~n=1,\dots,N,\label{eq:cons2}
\end{alignat}
where $R_k^{\rm min}$ is the minimum required rate of user $k$. Note that problem (P1) is always feasible thanks to the favorable propagation property shown in Theorem \ref{theorem0}. Specifically, given any feasible IRS reflecting coefficient solution satisfying (\ref{eq:cons2}), the target rate constraints (\ref{eq:cons1}) can be always satisfied with the following transmit power solution
\begin{align}
p_k=\frac{\left(2^{\frac{R_k^{\rm min}T}{T-K}}-1\right)\sigma^2}{M\beta^{\rm IU}_k\beta^{\rm BI}\mv{\phi}^H\bar{\mv{C}}_k\mv{\phi}}, ~~~ \forall k.
\end{align}

Next, constraint \eqref{eq:cons1} can be re-expressed as
\begin{align}
\frac{1}{p_k}\frac{\left(2^{\frac{R_k^{\rm min}T}{T-K}}-1\right)\sigma^2}{M\beta^{\rm IU}_k\beta^{\rm BI}}\le\mv{\phi}^H\bar{\mv{C}}_k\mv{\phi},~~k=1,\dots,K. \label{eq:cons1_new}
\end{align}
However, constraint \eqref{eq:cons1_new} is still non-convex since given $\bar{\mv{C}}_k\succ \mv{0}$, $\mv{\phi}^H\bar{\mv{C}}_k\mv{\phi}$ is a convex function, rather than a concave function, over $\mv{\phi}$, $\forall k$. In the following, we apply the successive convex approximation (SCA) technique to solve problem (P1) with constraint \eqref{eq:cons1} replaced by \eqref{eq:cons1_new}.

\subsection{Successive Convex Approximation Method}
For convenience, denote
\begin{align}
\mv{A}_k=\left[\begin{array}{ccccccc}
\mathfrak{Re}\left(\bar{\mv{C}}_k\right) &-\mathfrak{Im}\left(\bar{\mv{C}}_k\right) \\ \mathfrak{Im} \left(\bar{\mv{C}}_k\right) & \mathfrak{Re}\left(\bar{\mv{C}}_k\right)
\end{array}
\right],~~\forall k, \label{eq:A_k}
\end{align}
and
\begin{align}
\mv{b}=\left[\mathfrak{Re}\left(\mv{\phi}\right)^T ~~\mathfrak{Im}\left(\mv{\phi}\right)^T\right]^T, \label{eq:b_k}
\end{align}
where $\mathfrak{Re}(\mv{X})$ and $\mathfrak{Im}(\mv{X})$ denote the real and imaginary parts of $\mv{X}$, respectively. Since $\bar{\mv{C}}_k^H=\bar{\mv{C}}_k$, i.e., $\mathfrak{Re}\left(\bar{\mv{C}}_k\right)^T=\mathfrak{Re}\left(\bar{\mv{C}}_k\right)$ and $\mathfrak{Im}\left(\bar{\mv{C}}_k\right)^T=-\mathfrak{Im}\left(\bar{\mv{C}}_k\right)$, we have  $\mv{A}_k^T=\mv{A}_k$, $\forall k$. Then, it can be shown that $\mv{\phi}^H\bar{\mv{C}}_k\mv{\phi}$ can be characterized by the following real expression:
\begin{align}
\mv{\phi}^H\bar{\mv{C}}_k\mv{\phi}=\mv{b}^T\mv{A}_k\mv{b},~~\forall k.\label{eq:noncox}
\end{align}
Due to the convexity of $\mv{b}^T\mv{A}_k\mv{b}$ over $\mv{b}$, at any point $\bar{\mv{b}}$ that satisfies $\bar{\mv{b}}_n^2+\bar{\mv{b}}_{n+N}^2\le 1$, $\forall n$, where $\bar{\mv{b}}_{n}$ denotes the $n$-th element in $\bar{\mv{b}}$, its first-order Taylor approximation always serves as a lower bound:
\begin{align}
\mv{b}^T\mv{A}_k\mv{b}\ge\bar{\mv{b}}^T\mv{A}_k\bar{\mv{b}}+2\bar{\mv{b}}^T\mv{A}_k\left(\mv{b}-\bar{\mv{b}}\right), \forall k,\label{eq:app}
\end{align}
where the equality holds if and only if $\mv{b}=\bar{\mv{b}}$. Given any $\bar{\mv{b}}$ that satisfies $\bar{\mv{b}}_n^2+\bar{\mv{b}}_{n+N}^2\le 1$, $\forall n$, the approximated problem for problem (P1) is then given as
\begin{alignat}{3}
({\rm P2}):\quad
 &\min_{p_1,\dots,p_K,\mv{b}} &\quad &\sum\limits_{k=1}^{K} p_k \nonumber \\
&~~~\text{s.t.}                          &      & \frac{1}{p_k}\frac{\left(2^{\frac{R_k^{\rm min}T}{T-K}}-1\right)\sigma^2}{M\beta^{\rm IU}_k\beta^{\rm BI}}\le\bar{\mv{b}}^T\mv{A}_k\bar{\mv{b}}+2\bar{\mv{b}}^T\mv{A}_k\left(\mv{b}-\bar{\mv{b}}\right), ~~\forall k,\label{eq:app2}\\
&                                     &      & \mv{b}_n^2+\mv{b}_{n+N}^2\le 1, ~~n=1,\dots,N, \label{eq:cons3}
\end{alignat}
where $\mv{b}_n$ denotes the $n$-th element of $\mv{b}$. Problem (P2) is convex and can be efficiently solved by the existing convex optimization software, e.g., CVX\cite{grant2014cvx}.

\begin{algorithm}\label{alg1}
	\SetAlgoLined
	\textbf{Initialization}: For iteration $i=0$, generate initial $\bar{\mv{b}}$ satisfying $\bar{\mv{b}}_n^2+\bar{\mv{b}}_{n+N}^2\le 1$, $\forall n$;\\
	\textbf{Repeat}: \\
	\begin{enumerate}
		\item For iteration $i\ge1$, find the optimal solution to problem (P2) using CVX as $\mv{b}^{(i)}$ and $p_k^{(i)}$, $k=1,\dots,K$;
		\item Update $\bar{\mv{b}}=\mv{b}^{(i)}$ in \eqref{eq:app2};
		\item $i=i+1$;	
	\end{enumerate}
	\textbf{Until} $||\mv{b}^{(i)}-\mv{b}^{(i-1)}||_2^2\le\delta$.
	\caption{Proposed SCA Method to Solve Problem (P1)}
\end{algorithm}

After solving problem (P2) given the point $\bar{\mv{b}}$, the SCA method for problem (P1) proceeds by iteratively updating $\bar{\mv{b}}$ based on the solution to problem (P2). The proposed iterative algorithm is summarized in Algorithm \ref{alg1}, where $i$ denotes the index
of iteration, and $\delta>0$  is a given threshold. The convergence of Algorithm \ref{alg1} is guaranteed and the converged solution satisfies the Karush-Kuhn-Tucker (KKT) condition of the Problem (P1)\cite{marks1978general}.

\vspace{-5pt}
\subsection{Numerical Results}
In this subsection, we present numerical results to evaluate the performance the proposed SCA algorithm. We assume that the number of users is $K=4$. In addition, we use the channel covariance model 1 detailed in Section \ref{sec:protocol design} in the numerical examples. The other numerical parameters are the same as those used in Section \ref{sec:Date transmission}.

\begin{figure}[t]
	\centering
	\includegraphics[width=9cm]{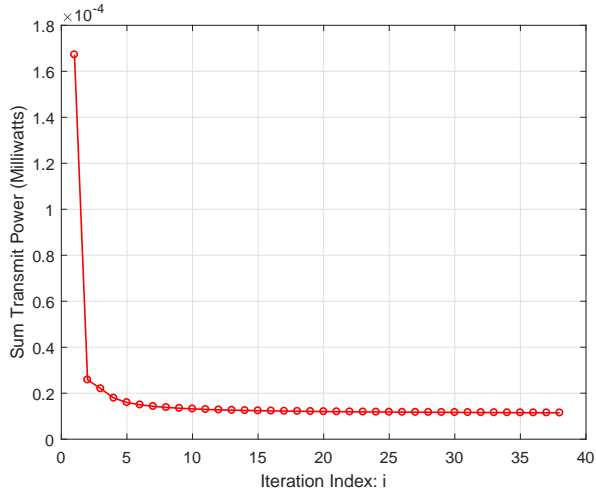}\vspace{-10pt}
	\caption{Convergence speed of Algorithm \ref{alg1} (SCA method) when $N=1280$ and $M=128$.}\label{CSpeed}\vspace{-15pt}
\end{figure}

We first show the convergence speed of Algorithm \ref{alg1} (i.e., the SCA algorithm) in Fig. \ref{CSpeed}. In this example, the target rates of the $4$ users are $R_1^{\rm min}=1$ bps/Hz, $R_2^{\rm min}=1.5$ bps/Hz, $R_3^{\rm min}=1.5$ bps/Hz, and $R_4^{\rm min}=2$ bps/Hz, respectively. As shown in Fig. \ref{CSpeed}, Algorithm \ref{alg1} converges very fast with iterations.

\begin{figure}[ht]
	\begin{center}
		\subfigure[$N=100$ and $M=5$]{\scalebox{0.55}{\label{Min_pw_compare2}\includegraphics*{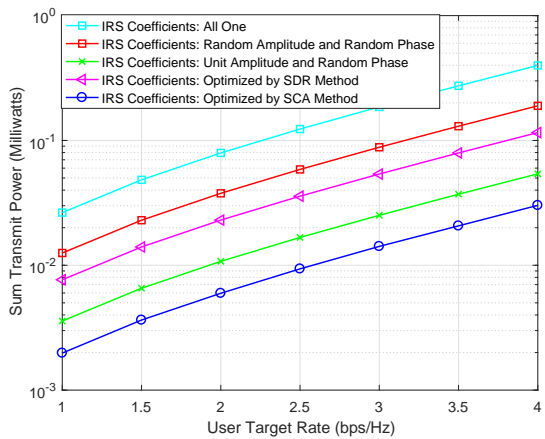}}}
		\subfigure[$N=1280$ and $M=128$.]{\scalebox{0.55}{\label{Min_pw_compare1}\includegraphics*{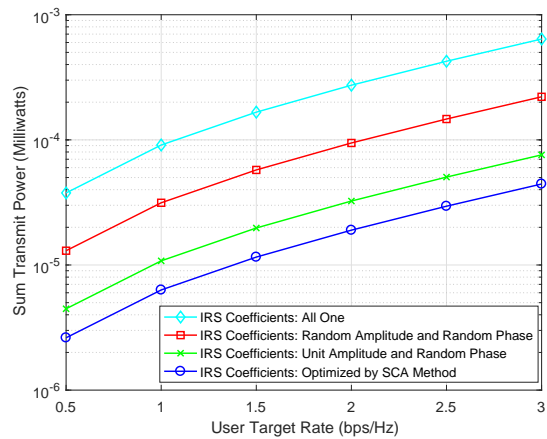}}}
	\end{center}
	\caption{\textbf{[Response 1.8]}  Performance comparison among different IRS reflection coefficients designs.}\label{Min_pw_compare}\vspace{-20pt}
\end{figure}
Second, we compare the performance of Algorithm \ref{alg1} (i.e., the SCA algorithm) with that of some benchmark schemes.
\begin{itemize}
	\item \textbf{IRS Coefficients: All One.} In this case, we set  $\phi_{n}=1$, $\forall n$.
	\item \textbf{IRS Coefficients: Random Amplitude and Random Phase.} In this case, the amplitude of each IRS reflection coefficient $\phi_n$ is randomly distributed between $(0 ~1]$, and the phase of each IRS reflection coefficient $\phi_n$ is randomly distributed between $(0 ~2\pi]$, $\forall n$.
    \item \textbf{IRS Coefficients: Unit Amplitude and Random Phase.} In this case, the amplitude of each IRS reflection coefficient is set to be one, and the phase of each IRS reflection coefficient is randomly distributed between $(0 ~2\pi]$, $\forall n$.
    \item \textbf{IRS Coefficients: Optimized by  SDR Method.} In this case, the IRS reflecting coefficients are optimized by the semidefinite relaxation (SDR) method \cite{wu2019intelligent}.
\end{itemize}

We first provide the performance comparison when $N=100$ and $M=5$ in  Fig. \ref{Min_pw_compare2} since the SDR method does not work with large $M$ and $N$ due to its complexity. The target rates are assumed to be the same for the all users. It is observed that the sum-power achieved by SCA algorithm is much smaller than that achieved by all the benchmark schemes. Next, we provide performance comparison when $N=1280$ and $M=128$ in Fig. \ref{Min_pw_compare1}, where the SDR method does not work. Again, it is observed that the sum-power achieved by SCA algorithm is much smaller than that achieved by the benchmark schemes.

\vspace{-5pt}
\section{Conclusions and Future Work}\label{sec:Conclusions}
In this paper, we proposed to deploy IRSs in massive MIMO systems to serve the users in the service dead zone. Specifically, to reduce the overhead of channel estimation and computational complexity of BS beamforming and IRS reflection design, we advocated a novel two-timescale protocol. Under the asymptotic regime where the numbers of BS antennas and IRS elements go to infinity with a fixed ratio, we first characterized the MMSE channel estimators of the user effective channels, then quantified the user achievable rates as functions of channel covariance matrices when the channel estimation overhead and error are taken into consideration, and last designed the IRS reflection coefficients to minimize the long-term transmit power of all the users based on channel covariance matrices. It was shown theoretically and numerically that the IRS can improve the performance of massive MIMO communication.

Some promising future work directions are described as follows. First, for IRS reflection coefficients design, this paper mainly considers the long-term power minimization problem. It is interesting to investigate how to optimize the other metrics based on our rate characterization results. Moreover, this paper assumes the Rayleigh fading model for the BS-IRS channels and the IRS-user channels. In practice, the IRS may be deployed closed to either the BS or the users such that the BS-IRS or IRS-user channels follow the Rician fading model. It is interesting to check whether the channel hardening and favorable propagation properties hold and characterize the user achievable rates under this channel model. Next, this paper mainly focuses on the issues arising from estimating the instantaneous CSI. It is also important to study the efficient way to estimate the statistical CSI, i.e., the channel covariance matrices, by utilizing some unique property of the IRS-assisted massive MIMO communication. Such a research line may lead to new theoretical problems about covariance estimation as well as practical solutions to implement the two-timescale communication protocol.

\begin{appendix}
\subsection{Useful Results}\label{appendix1}
\begin{lemma}[Lemma 14.2\cite{couillet2011random}] \label{lemma2}
Let $\mv{A}\in\mathbb{C}^{M\times M}$ be a random matrix with almost surely uniformly bounded spectral norm. Let $\mv{x}\in\mathbb{C}^{M\times 1}$ be a random vector with i.i.d. entries of zero mean, and variance $1/M$, and eighth order moment of order $O(1/N^4)$, and independent of $\mv{A}$. Then as $M\rightarrow \infty$,
\begin{align}
\mv{x}^H\mv{A}\mv{x}-\frac{1}{M}{\rm tr}(\mv{A})\xrightarrow{\text{a.s.}} 0.
\end{align}
\end{lemma}
\begin{lemma} \label{lemma2.5}
	Let $\mv{A}\in\mathbb{C}^{M\times M}$ be a random matrix with almost surely uniformly bounded spectral norm. Let $\mv{x}\in\mathbb{C}^{M\times 1}$ and $\mv{y}\in\mathbb{C}^{M\times 1}$ be mutually independent random vectors with i.i.d. entries of zero mean,  and variance $1/M$, and fourth order moment of order $O(1/N^2)$, and independent of $\mv{A}$. Then as $M\rightarrow \infty$,
	\begin{align}
	\mv{x}^H\mv{A}\mv{y}\xrightarrow{\text{a.s.}} 0.
	\end{align}
\end{lemma}
\begin{IEEEproof}
The proof is similar to that of Lemma 14.2 in \cite{couillet2011random}. We omit the details here.
\end{IEEEproof}

\subsection{Proof of Lemma \ref{lemma1}}\label{appendix4}
In the following, we first show that under Assumption \ref{assumption1.5}, all the elements in $\mv{q}$, i.e., $q_{k,m}$'s, $k=1,\ldots,K$ and $m=1,\ldots,M$, are Gaussian random variables when $N$ goes to infinity. Then, we show that $q_{k,m}$'s are independent over both $k$ and $m$ when $N$ goes to infinity. These two results are sufficient to prove Lemma \ref{lemma1}.

First, we show that all the elements in $\mv{q}$ are Gaussian. Denote the singular value decomposition (SVD) of  $(\mv{C}^{\rm I})^{\frac{1}{2}}{\rm diag}(\mv{\phi})(\mv{C}_k^{\rm I})^{\frac{1}{2}}$ within \eqref{eq: g} as
\begin{align}
	(\mv{C}^{\rm I})^{\frac{1}{2}}{\rm diag}(\mv{\phi})(\mv{C}_k^{\rm I})^{\frac{1}{2}}=\mv{U}_k\mv{\Lambda}_k\mv{G}_k^H,~~\forall k, \label{eq:svd}
\end{align}
where  $\mv{U}_k\in\mathbb{C}^{N\times N}$ and $\mv{G}_k\in\mathbb{C}^{N\times N}$ are unitary matrices, and $\mv{\Lambda}_k={\rm diag}(\sigma_{k,1},\dots,\sigma_{k,N})$ is a diagonal matrix whose diagonal elements are $N$ non-negative singular values. Substituting \eqref{eq:svd} into \eqref{eq: g}, we have
\begin{align}	\mv{q}_k=\frac{1}{\sqrt{N}}\bar{\mv{R}}_k\mv{\Lambda}_k\bar{\mv{t}}_k=\frac{1}{\sqrt{N}}\sum_{n=1}^{N}\bar{\mv{r}}_{k,n}\bar{t}_{k,n}\sigma_{k,n}=\frac{1}{\sqrt{N}}\sum_{n=1}^{N}\mv{m}_{k,n}\sigma_{k,n}, ~~\forall k, \label{eq: g_new2}
\end{align}
where $\bar{\mv{R}}_k=\tilde{\mv{R}}\mv{U}_k=\left[\bar{\mv{r}}_{k,1},\dots,\bar{\mv{r}}_{k,N}\right]$,  $\bar{\mv{t}}_k=\mv{G}_k^H\tilde{\mv{t}}_k=\left[\bar{t}_{k,1},\dots,\bar{t}_{k,N}\right]^T$, and  $\mv{m}_{k,n}=\bar{\mv{r}}_{k,n}\bar{t}_{k,n}$. Since $\mv{U}_k$ is a unitary matrix, the distribution of $\bar{\mv{R}}_k$ in (\ref{eq: g_new2}) is the same as that of $\tilde{\mv{R}}$, i.e.,  $\bar{\mv{R}}_k\sim \mathcal{CN}(\mv{0},N\beta^{{\rm BI}}\mv{I})$, $\forall k$. Similarly, since  $\mv{G}_k$ is a unitary matrix, the distribution of $\bar{\mv{t}}_k$ is the same as that of $\tilde{\mv{t}}_k$, i.e., $\bar{\mv{t}}_k\sim \mathcal{CN}(\mv{0}, \beta_k^{{\rm IU}}\mv{I})$, $\forall k$. In \eqref{eq: g_new2},  $\sigma_{k,n}$'s are fixed terms given $\mv{C}^{\rm I}$, $\mv{C}_k^{\rm I}$'s, and $\mv{\phi}$, and   $\mv{m}_{k,n}$'s are independent over $n$ since $\bar{\mv{r}}_{k,n}$'s and $\bar{t}_{k,n}$'s are independent over $n$, respectively. Moreover,  according to \eqref{eq:lk} in Assumption \ref{assumption1.5},   $\sigma_{k,n}>l_k>0$, $\forall n$, where the lower bound $l_k$ is independent of $N$. Thus, \eqref{eq: g_new2} shows that $\mv{q}_k$ is the summation of $N$ independent vectors.  From the Multivariate Lindeberg-Feller Central Limit theorem in \cite{van2000asymptotic}, we can approximate $\mv{q}_k$ as a Gaussian random vector when $N$ goes to infinity, $k=1,\dots,K$. In other words, all the elements in $\mv{q}$ are Gaussian.

Next, we show that $q_{k,m}$'s are independent over both $k$ and $m$ when $N$ goes to infinity. First, it can be shown that within a vector $\mv{q}_k$, $E[q_{k,m_1}q_{k,m_2}^\ast]=0$ holds $\forall m_1\neq m_2$, i.e., $q_{k,m_1}$ and $q_{k,m_2}$ are uncorrelated. Since $\mv{q}_k$ is a Gaussian vector, it follows that for any given $k$, $q_{k,m}$'s are independent over $m$. Then, we aim to show that $\forall k_1\neq k_2$: 1. $q_{k_1,m_1}$ and $q_{k_2,m_2}$ are independent with each other $\forall m_1\neq m_2$; and 2. $q_{k_1,m}$ and $q_{k_2,m}$ are independent with each other.

Consider the case when $m_1\neq m_2$. According to \eqref{eq: g_new2}, we have
\begin{align} q_{k,m}=\frac{1}{\sqrt{N}}\sum_{n=1}^{N}\bar{r}_{k,n,m}\bar{t}_{k,n}\sigma_{k,n}=\tilde{\mv{r}}_m^T\frac{1}{\sqrt{N}}\sum_{n=1}^{N}\mv{u}_{k,n}\bar{t}_{k,n}\sigma_{k,n}, ~ \forall k,m, \label{eq: g_new2_1}
\end{align}
where $\bar{r}_{k,n,m}$ is the $m$-th element of $\bar{\mv{r}}_{k,n}$ in \eqref{eq: g_new2}, $\tilde{\mv{r}}_m^T$ is the $m$-th row of $\tilde{\mv{R}}$ in \eqref{eq: R}, and $\mv{u}_{k,n}$ is the  $n$-th column of $\mv{U}_k$ in \eqref{eq:svd}. Then, according to \eqref{eq: g_new2_1}, given any $m_1\ne m_2$ and $k_1\ne k_2$, it follows that $\tilde{\mv{r}}_{m_1}$ is independent with $\tilde{\mv{r}}_{m_2}$ and $\bar{t}_{k_2,n}$'s, $\forall n$, and $\bar{t}_{k_1,n}$'s, $\forall n$, are independent with $\tilde{\mv{r}}_{m_2}$ and $\bar{t}_{k_2,n}$'s, $\forall n$. As a result, (\ref{eq: g_new2_1}) implies that $q_{k_1,m_1}$ and $q_{k_2,m_2}$ are independent when $k_1\neq k_2$ and $m_1\neq m_2$.

Next, we consider the case when $m_1=m_2=m$ and show that $q_{k_1,m}$ and $q_{k_2,m}$ are independent with each other when $k_1\neq k_2$. According to \cite{pishro2016introduction}, we first aim to prove that $q_{k_1,m}$ and $q_{k_2,m}$ are jointly Gaussian by showing that for arbitrary $a_{k_1}$ and $a_{k_2}$, $b_{m,k_1,k_2}= a_{k_1}q_{k_1,m}+a_{k_2}q_{k_2,m}$ is a Gaussian random variable when $N$ goes to infinity. If both $a_{k_1}$ and $a_{k,2}$ are zero, then $b_{m,k_1,k_2}=0$ is a Gaussian random variable. Otherwise, we can express $b_{m,k_1,k_2}$ as
\begin{align}
	b_{m,k_1,k_2}=a_{k_1}q_{k_1,m}+a_{k_2}q_{k_2,m}=\frac{1}{\sqrt{N}}\tilde{\mv{r}}_m^T(\mv{C}^{\rm I})^{\frac{1}{2}}{\rm diag}(\mv{\phi})(a_{k_1}(\mv{C}_{k_1}^{\rm I})^{\frac{1}{2}}\tilde{\mv{t}}_{k_1}+a_{k_2}(\mv{C}_{k_2}^{\rm I})^{\frac{1}{2}}\tilde{\mv{t}}_{k_2}).\label{eq:b}
\end{align}Since $\tilde{\mv{t}}_{k_1}$ and $\tilde{\mv{t}}_{k_2}$ are independent Gaussian vectors, it follows that $b_{m,k_1,k_2}$ has the same distribution as the following random variable:
\begin{align}
	\bar{b}_{m,k_1,k_2}=\frac{1}{\sqrt{N}}\tilde{\mv{r}}_m^T\mv{C}_{k_1,k_2}\mv{t}_{k_1,k_2},\label{eq:b1}
\end{align}where $\mv{C}_{k_1,k_2}=(\mv{C}^{\rm I})^{\frac{1}{2}}{\rm diag}(\mv{\phi})(\beta_{k_1}^{{\rm IU}}a_{k_1}^2\mv{C}_{k_1}^{\rm I}+\beta_{k_2}^{{\rm IU}}a_{k_2}^2\mv{C}_{k_2}^{\rm I})^{\frac{1}{2}}$, and   $\mv{t}_{k_1,k_2}\in\mathbb{C}^{N\times 1}\sim \mathcal{CN}(\mv{0}, \mv{I})$. Denote the SVD of  $\mv{C}_{k_1,k_2}$ in \eqref{eq:b1} as
\begin{align}
	\mv{C}_{k_1,k_2}=\bar{\mv{U}}_{k_1,k_2}\bar{\mv{\Lambda}}_{k_1,k_2}\bar{\mv{G}}_{k_1,k_2}^H, \label{eq:svd3}
\end{align}
where  $\bar{\mv{U}}_{k_1,k_2}\in\mathbb{C}^{N\times N}$ and $\bar{\mv{G}}_{k_1,k_2}\in\mathbb{C}^{N\times N}$ are unitary matrices, and $\bar{\mv{\Lambda}}_{k_1,k_2}={\rm diag}(\bar{\sigma}_{k_1,k_2,1},\dots,\bar{\sigma}_{k_1,k_2,N})$ is a diagonal matrix whose diagonal elements consist of the $N$ non-negative singular values of $\mv{C}_{k_1,k_2}$. Then, according to Weyl's theorem in \cite{horn2012matrix}, under Assumption \ref{assumption1.5}, we have
\begin{align}
	\bar{\sigma}_{k_1,k_2,n}\ge\sqrt{\beta_{k_1}^{{\rm IU}}a_{k_1}^2\left(\sigma_{k_1}^{\rm min}\right)^2+\beta_{k_2}^{{\rm IU}}a_{k_2}^2\left(\sigma_{k_2}^{\rm min}\right)^2}>\sqrt{\beta_{k_1}^{{\rm IU}}a_{k_1}^2l_{k_1}^2+\beta_{k_2}^{{\rm IU}}a_{k_2}^2l_{k_2}^2}, ~~~ \forall n.\label{eq: simg}
\end{align}Therefore, if at least one of $a_{k_1}$ and $a_{k_2}$ is non-zero, each $\bar{\sigma}_{k_1,k_2,n}$, $n=1,\ldots,N$, is lower bounded by a value that is independent of $N$. Substituting \eqref{eq:svd3} into \eqref{eq:b1}, we have
\begin{align}	b_{m,k_1,k_2}=\frac{1}{\sqrt{N}}\hat{\mv{r}}_{m,k_1,k_2}^T\bar{\mv{\Lambda}}_{k_1,k_2}\hat{\mv{t}}_{k_1,k_2}=\frac{1}{\sqrt{N}}\sum_{n=1}^N\bar{\sigma}_{n,k_1,k_2}\hat{r}_{m,k_1,k_2,n}\hat{t}_{k_1,k_2,n},\label{eq:b2}
\end{align}
where $\hat{\mv{r}}_{m,k_1,k_2}^T=\tilde{\mv{r}}_m^T\bar{\mv{U}}_{k_1,k_2}\sim \mathcal{CN}(\mv{0}, \beta^{{\rm BI}}\mv{I})$ with $\hat{r}_{m,k_1,k_2,n}$ being its $n$-th element, and  $\hat{\mv{t}}_{k_1,k_2}=\bar{\mv{G}}_{k_1,k_2}^H\mv{t}_{k_1,k_2}\sim \mathcal{CN}(\mv{0}, \mv{I})$ with $\hat{t}_{k_1,k_2,n}$ being its $n$-th element. Since $\bar{\sigma}_{k_1,k_2,n}$ is lower bounded by a value that is independent of $N$ according to \eqref{eq: simg}, $\forall n$, $\bar{b}_{m,k_1,k_2}$ is the summation of $N$ independent random variables. According to the Central Limit theorem in \cite{van2000asymptotic},  $\bar{b}_{m,k_1,k_2}$ and also $b_{m,k_1,k_2}$ are Gaussian random variables when $N$ goes to infinity. By combining the cases that both $a_{k_1}$ and $a_{k_2}$ are zero and at least one of $a_{k_1}$ and $a_{k_2}$ is non-zero, it follows that $q_{k_1,m}$ and $q_{k_2,m}$ are jointly Gaussian. Moreover, the correlation between $q_{k_1,m}$ and $q_{k_2,m}$ is
\begin{align}	\mathbb{E}\left(q_{k_1,m}q_{k_2,m}^\ast\right)=\frac{1}{{N}}\mathbb{E}\left[\sum_{n=1}^{N}\bar{r}_{k_1,n,m}\bar{t}_{k_1,n}\sigma_{k_1,n}\left(\sum_{n=1}^{N}\bar{r}_{k_2,n,m}\bar{t}_{k_2,n}\sigma_{k_2,n}\right)^\ast\right]=0,
\end{align}
since $\bar{t}_{k,n}$'s are mutually independent over $n$ and $k$. Because $q_{k_1,m}$ and $q_{k_2,m}$ are jointly Gaussian and uncorrelated, they are also independent.

Until now, we have shown that for any $k$, $q_{k,m_1}$ and $q_{k,m_2}$ are independent if $m_1\neq m_2$; for any $k_1\neq k_2$, $q_{k_1,m_1}$ and $q_{k_2,m_2}$ are independent under both the cases that $m_1\neq m_2$ and $m_1=m_2$. As a result, all the elements in $\mv{q}=\left[\mv{q}_1^T,\dots,\mv{q}_K^T\right]^T$ are independent. Moreover, all the elements in $\mv{q}$ are Gaussian. As a result, $\mv{q}$ is a Gaussian vector. Lemma \ref{lemma1} is thus proved.

\subsection{Proof of Theorem \ref{theorem0}}  \label{appendix2}
We first prove the asymptotic behavior of $\frac{\mv{c}_k^H\mv{c}_j}{MN}$ shown in \eqref{eq: chan-fav}. As $M,N$ go to infinity with $N/M=q$,
\begin{align}
&\frac{\mv{c}_k^H\mv{c}_j}{MN}=q\frac{\mv{c}_k^H\mv{c}_j}{N^2}=q\frac{\left(\mv{h}_k+\sum_{n=1}^{N}\phi_{n}\mv{g}_{k,n}\right)^H\left(\mv{h}_j+\sum_{n=1}^{N}\phi_{n}\mv{g}_{j,n}\right)}{N^2} \nonumber\\
&=q\frac{\tilde{\mv{h}}_k^H(\mv{C}_k^{\rm B})^{\frac{1}{2}}(\mv{C}_j^{\rm B})^{\frac{1}{2}}\tilde{\mv{h}}_j}{N^2}
+q\frac{\tilde{\mv{h}}_k^H\hat{\mv{C}}_{k,j}\tilde{\mv{t}}_j}{N}+q\frac{(\tilde{\mv{t}}_k)^H\hat{\mv{C}}_{k,j}^H\tilde{\mv{h}}_j}{N}+q\frac{(\tilde{\mv{t}}_k)^H\tilde{\mv{C}}_{k,j}\tilde{\mv{t}}_j}{N}, \label{eq: III-A2.1}
\end{align}
where $\tilde{\mv{C}}_{k,j}$ and $\hat{\mv{C}}_{k,j}$ are shown in \eqref{eq: III-A0} and \eqref{eq: III-A01}, respectively.
In \eqref{eq: III-A2.1}, since $\mv{C}_k^{\rm B}$, $\forall k$, have uniformly bounded spectral norm, $(\mv{C}_k^{\rm B})^{\frac{1}{2}}(\mv{C}_j^{\rm B})^{\frac{1}{2}}$, $\forall k\ne j$,  have uniformly bounded spectral norm. In addition, $\tilde{\mv{h}}_k$ and $\tilde{\mv{h}}_j$ are mutually independent vectors with i.i.d. entries, $\forall k\ne j$. Thus, according to Theorem 3.7 in \cite{couillet2011random}, as $M,N$ go to infinity with $N/M=q$, we have
\begin{align}
q\frac{\tilde{\mv{h}}_k^H(\mv{C}_k^{\rm B})^{\frac{1}{2}}(\mv{C}_j^{\rm B})^{\frac{1}{2}}\tilde{\mv{h}}_j}{N^2}\xrightarrow{\text{a.s.}}0,~\forall k\ne j. \label{eq: III-A2.5}
\end{align}
Moreover, denote the SVD of $\hat{\mv{C}}_{k,j}$ as
\begin{align}
\hat{\mv{C}}_{k,j}=\mv{L}_{k,j}\mv{\Sigma}_{k,j}\mv{O}_{k,j}^H,~~\forall k\ne j, \label{eq:svd2}
\end{align}
where  $\mv{L}_{k,j}\in\mathbb{C}^{M\times M}$ and $\mv{O}_{k,j}\in\mathbb{C}^{N\times N}$ are unitary matrices. First, consider the case that $N\geq M$. In this case, $\mv{\Sigma}_{k,j}\in\mathbb{C}^{M\times N}$ is expressed as
\begin{align}
\mv{\Sigma}_{k,j}=\left[
\mv{\Sigma}_{k,j}^{(1)} ~~\mv{0}
\right], \label{eq:sigma}
\end{align}
where $\mv{\Sigma}_{k,j}^{(1)}={\rm diag}(\delta_{k,j}^{(1)},\dots,\delta_{k,j}^{(M)})$ contains the $M$ positive singular values of $\hat{\mv{C}}_{k,j}$. Then,
\begin{align}
q\frac{\tilde{\mv{h}}_k^H\hat{\mv{C}}_{k,j}\tilde{\mv{t}}_j}{N}=q\frac{\hat{\mv{h}}_{k,j}^H\mv{\Sigma}_{k,j}\hat{\mv{t}}_{k,j}}{N}=q\frac{\hat{\mv{h}}_{k,j}^H\mv{\Sigma}_{k,j}^{(1)}\hat{\mv{t}}_{k,j}^{(1)}}{N},
\end{align}
where
$\hat{\mv{h}}_{k,j}=\mv{L}_{k,j}^H\tilde{\mv{h}}_k$ and $\hat{\mv{t}}_{k,j}=\mv{O}_{k,j}^H\tilde{\mv{t}}_j=\left[\left(\hat{\mv{t}}_{k,j}^{(1)}\right)^T~\left(\hat{\mv{t}}_{k,j}^{(2)}\right)^T\right]^T$ with $\hat{\mv{t}}_{k,j}^{(1)}\in\mathbb{C}^{M\times 1}$ and  $\hat{\mv{t}}_{k,j}^{(2)}\in\mathbb{C}^{(N-M)\times 1}$. Since $\mv{L}_{k,j}$ is a unitary matrix, the distribution of $\hat{\mv{h}}_{k,j}$ is the same as that of $\tilde{\mv{h}}_k$, i.e.,  $\hat{\mv{h}}_{k,j}\sim \mathcal{CN}(\mv{0},\beta_k^{{\rm BU}}\mv{I})$, $\forall k$. Similarly, it follows that $\hat{\mv{t}}_{k,j}\sim \mathcal{CN}(\mv{0}, \beta_k^{{\rm IU}}\mv{I})$, $\forall k$. Since  $\mv{\Sigma}_{k,j}^{(1)}$ has almost surely uniformly bounded spectral norm according to Assumption \ref{assumption2}, by utilizing  Lemma \ref{lemma2.5} in Appendix \ref{appendix1} we have
\begin{align}
q\frac{\tilde{\mv{h}}_k^H\hat{\mv{C}}_{k,j}\tilde{\mv{t}}_j}{N}\xrightarrow{\text{a.s.}}0.\label{eq: III-A2.6}
\end{align}
We can also get the same results when $M>N$.
Similarly, for the third and fourth terms in \eqref{eq: III-A2.1},
\begin{align}
& q\frac{(\tilde{\mv{t}}_k)^H\hat{\mv{C}}_{k,j}^H\tilde{\mv{h}}_j}{N}\xrightarrow{\text{a.s.}}0,\label{eq: III-A2.65} \\
& q\frac{(\tilde{\mv{t}}_k)^H\tilde{\mv{C}}_{k,j}\tilde{\mv{t}}_j}{N}\xrightarrow{\text{a.s.}}0. \label{eq: III-A2.7}
\end{align}
Substituting \eqref{eq: III-A2.5}, \eqref{eq: III-A2.6}, \eqref{eq: III-A2.65}, and \eqref{eq: III-A2.7} into \eqref{eq: III-A2.1}, as $M,N$ go to infinity with $N/M=q$, we have
\begin{align}
\frac{\mv{c}_k^H\mv{c}_j}{MN}\xrightarrow{\text{a.s.}}0, ~\forall k\ne j. \label{eq:phi2}
\end{align}

Then, we  show the asymptotic behavior of $\frac{\mv{c}_k^H\mv{c}_k}{MN}$ shown in \eqref{eq: chan-fav}. As  $M,N$ go to infinity with $N/M=q$,  we have
\begin{align}
\frac{\mv{c}_k^H\mv{c}_k}{MN}&=q\frac{\mv{c}_k^H\mv{c}_k}{N^2}=q\frac{\left(\mv{h}_k+\sum_{n=1}^{N}\phi_{n}\mv{g}_{k,n}\right)^H\left(\mv{h}_k+\sum_{n=1}^{N}\phi_{n}\mv{g}_{k,n}\right)}{N^2}\\&\xrightarrow{\text{a.s.}}q\frac{\tilde{\mv{h}}_k^H\mv{C}_k^{\rm B}\tilde{\mv{h}}_k}{N^2}+q\frac{(\tilde{\mv{t}}_k)^H\tilde{\mv{C}}_{k,k}\tilde{\mv{t}}_k}{N}. \label{eq: A9}
\end{align}
In \eqref{eq: A9},  since $\mv{C}_k^{\rm B}$ has uniformly bounded spectral norm and $\tilde{\mv{h}}_k$ has i.i.d. distributed random entries, according to Theorem 3.4 in \cite{couillet2011random}, as $M,N$ go to infinity with $N/M=q$, we have
\begin{align}
q\frac{\tilde{\mv{h}}_k^H\mv{C}_k^{\rm B}\tilde{\mv{h}}_k}{N^2}\xrightarrow{\text{a.s.}}q\frac{\beta_k^{\rm BU}{\rm tr}(\mv{C}_k^{\rm B})}{N^2}=\frac{\beta_k^{\rm BU}}{N}\xrightarrow{\text{a.s.}}0. \label{eq: III-A5}
\end{align}
In addition,
since  $\tilde{\mv{C}}_{k,k}$ has almost surely uniformly bounded
spectral norm based on Assumption \ref{assumption2} and $\tilde{\mv{t}}_k$ has i.i.d. random entries, according to Lemma \ref{lemma2.5} in Appendix \ref{appendix1}, as $M,N$ go to infinity with $N/M=q$, we have
\begin{align}
q\frac{(\tilde{\mv{t}}_k)^H\tilde{\mv{C}}_{k,k}\tilde{\mv{t}}_k}{N}&\xrightarrow{\text{a.s.}}q\beta^{\rm IU}_k\frac{{\rm tr}(\tilde{\mv{C}}_{k,k})}{N}=q\beta^{\rm IU}_k\frac{{\rm tr}\left(\tilde{\mv{R}}^H\mv{C}^{\rm B}\tilde{\mv{R}}\mv{D}_k(\mv{\phi})\right)}{N^2}.\label{eq: III-A8.1}
\end{align}
Denote the eigenvalue decomposition  of  $\mv{C}^{\rm B}$  as
\begin{align}
\mv{C}^{\rm B}=\mv{W}\mv{\Psi}\mv{W}^H, \label{eq:evd}
\end{align}
where  $\mv{W}\in\mathbb{C}^{M\times M}$ is a unitary matrix, and $\mv{\Psi}={\rm diag}(\psi_{1},\dots,\psi_{M})$ is a diagonal matrix whose diagonal elements are $M$ eigenvalues. Substituting \eqref{eq:evd} into \eqref{eq: III-A8.1}, we have
\begin{align}
&q\frac{(\tilde{\mv{t}}_k)^H\tilde{\mv{C}}_{k,k}\tilde{\mv{t}}_k}{N}\xrightarrow{\text{a.s.}}q\beta^{\rm IU}_k\frac{{\rm tr}\left(\hat{\mv{R}}\mv{\Psi}\hat{\mv{R}}^H\mv{D}_k(\mv{\phi})\right)}{N^2}=q\beta^{\rm IU}_k\frac{{\rm tr}\left(\sum_{m=1}^{M}\psi_m\hat{\mv{r}}_m\hat{\mv{r}}_m^H\mv{D}_k(\mv{\phi})\right)}{N^2}\nonumber\\
&=q\beta^{\rm IU}_k\sum_{m=1}^{M}\psi_m\frac{\hat{\mv{r}}_m^H\mv{D}_k(\mv{\phi})\hat{\mv{r}}_m}{N^2}\label{eq: III-A8.2}
\end{align}
where $\hat{\mv{R}}=\tilde{\mv{R}}^H\mv{W}=\left[\hat{\mv{r}}_1,\dots,\hat{\mv{r}}_M\right]$. In \eqref{eq: III-A8.2}, since the spectral norm
\begin{align}
||\mv{D}_k(\mv{\phi})||_2\le||\mv{C}^{\rm I}||_2||\mv{C}^{\rm I}_k||_2,
\end{align}
according to Assumption \ref{assumption1}, $\mv{D}_k(\mv{\phi})$ has uniformly bounded spectral norm. In this case, based on Theorem 3.4 in \cite{couillet2011random}, \eqref{eq: III-A8.2} can be further expressed as
\begin{align}
&q\frac{(\tilde{\mv{t}}_k)^H\tilde{\mv{C}}_{k,k}\tilde{\mv{t}}_k}{N}\xrightarrow{\text{a.s.}}q\beta^{\rm IU}_k\beta^{\rm BI}\frac{\sum_{m=1}^{M}\psi_m{\rm tr}\left(\mv{D}_k(\mv{\phi})\right)}{N^2}=q\beta^{\rm IU}_k\beta^{\rm BI}\frac{{\rm tr}\left(\mv{C}^{\rm B}\right){\rm tr}\left(\mv{D}_k(\mv{\phi})\right)}{N^2}\label{eq: III-A8.3}\\
&=\beta^{\rm IU}_k\beta^{\rm BI}\frac{{\rm tr}\left(\mv{D}_k(\mv{\phi})\right)}{N},\label{eq: III-A3}
\end{align}
where \eqref{eq: III-A8.3} to \eqref{eq: III-A3} follows from the fact that ${\rm tr}(\mv{C}^{\rm B})=M$ when $\mv{C}^{\rm B}$ is a channel correlation matrix with all diagonal elements being one. Substituting \eqref{eq: III-A5} and \eqref{eq: III-A3} into \eqref{eq: A9}, we have
\begin{align}
\frac{\mv{c}_k^H\mv{c}_j}{MN}\xrightarrow{\text{a.s.}}\frac{\beta^{\rm IU}_k\beta^{\rm BI}{\rm tr}\left(\mv{D}_k(\mv{\phi})\right)}{N}=\frac{\beta^{\rm IU}_k\beta^{\rm BI}\mv{\phi}^H\bar{\mv{C}}_k\mv{\phi}}{N},~\forall k,\label{eq: Phi1}
\end{align}
where $\bar{\mv{C}}_k$ is shown in \eqref{eq: C_k}. That is,
\begin{align}
\frac{\mv{c}_k^H\mv{c}_j}{MN}-\frac{\beta^{\rm IU}_k\beta^{\rm BI}\mv{\phi}^H\bar{\mv{C}}_k\mv{\phi}}{N}\xrightarrow{\text{a.s.}}0,~\forall k.\label{eq: Phi2}
\end{align}Theorem \ref{theorem0} is thus proved.

\subsection{Proof of Theorem \ref{theorem1}}\label{appendix3}
Substituting $p_k=\frac{E_k}{MN}$ into \eqref{eq:Sys-A1.7},  we have
\begin{align}
&\gamma_k(\mv{\phi})=\frac{E_k\left|\hat{\mv{c}}_k^H\hat{\mv{c}}_k/M\right|^2}{\sum_{j\ne k}^KE_j\left|\hat{\mv{c}}_k^H\hat{\mv{c}}_j/M\right|^2+\sum_{j=1}^{K}E_j\hat{\mv{c}}_k^H\mv{F}_j\hat{\mv{c}}_k/M^2+\sigma^2\hat{\mv{c}}_k^H\hat{\mv{c}}_k/M},~~\forall k. \label{eq:Sys-gamma_inf}
\end{align}
In \eqref{eq:Sys-gamma_inf},
we have already shown the asymptotic behaviors of $\frac{\hat{\mv{c}}_k^H\hat{\mv{c}}_k}{M}$ and $\frac{\hat{\mv{c}}_k^H\hat{\mv{c}}_j}{M}$  in \eqref{eq:phi21} and \eqref{eq:phi1}, respectively. In the following, we show the asymptotic behavior of $\frac{\mv{\hat{c}}_k^H\mv{F}_j\mv{\hat{c}}_k}{M^2}$.

From \eqref{eq:Fk_de}, we have
\begin{align}
\mv{F}_j=\frac{1}{N}\mv{B}_j^H{\rm diag}\left(\frac{\lambda_{j,1}\sigma^2}{\sigma^2+p^tK\lambda_{j,1}},\dots,\frac{\lambda_{j,M}\sigma^2}{\sigma^2+p^tK\lambda_{j,M}}\right)\mv{B}_j, \forall j.
\end{align}
Then,
\begin{align}
	\mv{\hat{c}}_k^H\mv{F}_j\mv{\hat{c}}_k=\frac{1}{N}\mv{\hat{c}}_k^H\mv{B}_j^H{\rm diag}\left(\frac{\lambda_{j,1}\sigma^2}{\sigma^2+p^tK\lambda_{j,1}},\dots,\frac{\lambda_{j,M}\sigma^2}{\sigma^2+p^tK\lambda_{j,M}}\right)\mv{B}_j\mv{\hat{c}}_k\le \frac{\sigma^2}{Np^tK}\mv{\hat{c}}_k^H\mv{\hat{c}}_k.
\end{align}
As $M$, $N$ go to infinity with $N/M=q$,
\begin{align}
0\le\frac{\mv{\hat{c}}_k^H\mv{F}_j\mv{\hat{c}}_k}{M^2}\le \frac{\sigma^2}{p^tK}\frac{\mv{\hat{c}}_k^H\mv{\hat{c}}_k}{M^2N}\xrightarrow{\text{a.s.}}0,
\end{align}
where the last equality follows from the fact that the power of $\mv{\hat{c}}_k$ is in the order of $M$. Thus, as $M$, $N$ go to infinity with $N/M=q$,
\begin{align}
	\frac{\mv{\hat{c}}_k^H\mv{F}_j\mv{\hat{c}}_k}{M^2}\xrightarrow{\text{a.s.}}0.\label{eq: III-B8}
\end{align}
Substituting $\frac{\hat{\mv{c}}_k^H\hat{\mv{c}}_k}{M}$ in \eqref{eq:phi21}, $\frac{\hat{\mv{c}}_k^H\hat{\mv{c}}_j}{M}$ in \eqref{eq:phi1}, and $\frac{\hat{\mv{c}}_k^H\mv{F}_j\hat{\mv{c}}_k}{M^2}$ in \eqref{eq: III-B8} into \eqref{eq:Sys-gamma_inf}, as $M$, $N$ go to infinity with $N/M=q$, we have
\begin{align}
\gamma_k(\mv{\phi})-\bar{\gamma_k}(\mv{\phi})\xrightarrow{\text{a.s.}}0,~\forall k, \label{eq:III-B81}
\end{align}
where
\begin{align}
\bar{\gamma_k}(\mv{\phi})=\beta^{\rm IU}_k\beta^{\rm BI}\frac{E_k\mv{\phi}^H\bar{\mv{C}}_k\mv{\phi}}{N\sigma^2}. \label{eq:III-B8}
\end{align}
From \eqref{eq:Sys-1.6}, Theorem \ref{theorem1} is thus proved.

\end{appendix}

\bibliographystyle{IEEEtran}
\bibliography{CIC}

\end{document}